\documentclass[english,13pt]{article}
\usepackage[T1]{fontenc}
\usepackage[latin1]{inputenc}
\usepackage{geometry}
\usepackage{float}
\usepackage{mathrsfs}
\usepackage{amsmath}
\usepackage{amssymb}
\usepackage{graphicx}
\usepackage[normalem]{ulem}
\usepackage{hyperref}
\hypersetup{
    colorlinks=true,
    linkcolor=blue,
    filecolor=magenta,      
    urlcolor=cyan,
    citecolor = red,
}

\usepackage{booktabs}

\makeatletter

\usepackage{algorithm,algpseudocode}

\usepackage{amsthm}

\usepackage{mathrsfs}

\usepackage{amsfonts}

\usepackage{epsfig}

\usepackage{bm}

\usepackage{mathrsfs}

\usepackage{enumerate}

\numberwithin{equation}{section}

\@ifundefined{definecolor}{\@ifundefined{definecolor}
 {\@ifundefined{definecolor}
 {\usepackage{color}}{}
}{}
}{}

\renewcommand{\textcolor}[2]{#2}

\usepackage{subfig}\usepackage[all]{xy}

\newtheorem{rem}{Remark}[section]
\newtheorem{prop}{Proposition}[section]

\newcounter{hypA}
\newenvironment{hypA}{\refstepcounter{hypA}\begin{itemize}
  \item[({\bf A\arabic{hypA}})]}{\end{itemize}}
\newcounter{hypB}

\newcounter{hypD}

\newcounter{hypW}

\usepackage{babel}\date{}

\usepackage{babel}

\makeatother

\usepackage{babel}

\begin{document}

\begin{center}

{\Large \textbf{Particle Filtering for a Class of State-Space Models with Low and Degenerate Observational Noise}}

\vspace{0.5cm}

BY  ABYLAY ZHUMEKENOV$^{1}$,  ALEXANDROS BESKOS$^{2}$,  DAN CRISAN$^{3}$, AJAY JASRA$^{1}$ \& NIKOLAS KANTAS$^{3}$

{\footnotesize $^{1}$School of Data Science,  The Chinese University of Hong Kong,  Shenzhen,  Shenzhen, CN.}\\
{\footnotesize $^{2}$Department of Statistical Science, University College London, London, WC1E 6BT, UK}\\
{\footnotesize $^{3}$Department of Mathematics, Imperial College London, London, SW7 2AZ, UK}\\
{\footnotesize E-Mail:\,} \texttt{\emph{\footnotesize  abylayzhumekenov@cuhk.edu.cn; 
a.beskos@ucl.ac.uk; d.crisan@ic.ac.uk;  
ajayjasra@cuhk.edu.cn; n.kantas@ic.ac.uk
}}

\end{center}

\begin{abstract}
We consider the discrete-time filtering problem in scenarios where the observation noise is low or degenerate. We focus on the case where the observation equation is a linear function of the state and the data involve additive noise. However, we place minimal assumptions on the hidden state process. For such a class of models we derive new particle filters (PFs) with the key property that their performance is robust to the size of the observation noise. As a consequence, the developed PFs are well-defined in the limiting case of degenerate observation noise. Indicatively, we prove (under assumptions) that the PF applied in this low noise setting inherits the properties of the PF used in the degenerate case. We extend our framework to the case where the hidden states are drawn from a diffusion process. 
We illustrate our algorithms numerically on several examples.
\end{abstract}

\section{Introduction}

The discrete-time filtering problem considers an unobserved Markov chain $\{X_n\}_{n\geq 0}$ (the signal or hidden chain) and associated data
$\{Y_n\}_{n\geq 1}$.  The data at any time $k$ are assumed to depend on   $\{X_n\}_{n\geq 0}$ only via the position $X_k$ of the hidden state.  The objective of filtering is to compute expectations with respect to (w.r.t.) the sequence of \emph{filtering distributions}, i.e.~the conditional distribution of the signal at each time $n$,  given all the data observed up to that time, sequentially in the time index. 
The filtering problem has numerous applications in statistics, applied mathematics, finance and beyond;  we refer the reader to the texts \cite{bain,cappe,delmoral} and the references therein for a more complete summary.

In most cases of practical interest  the filter is unavailable analytically and one must resort to numerical approximations.  A plethora of relevant algorithms have been developed,  see e.g.~\cite{bain,cappe,delmoral} for some coverage of the subject.  In this article we focus on the approach called particle filters (PFs).  These are Monte Carlo,  simulation-based methods that simulate a collection of $N\in\mathbb{N}$ samples in parallel and that undergo importance sampling and resampling operations. Such methods are very well understood from a mathematical perspective (e.g.~\cite{delmoral}) and provide almost surely consistent approximations of the filter as $N\rightarrow\infty$.  PFs are typically effective when the dimension of the hidden state, $d_x$,  is moderate,  although there are several recent attempts to further extend the methodology to cover some special-case high-dimensional problems (e.g.~\cite{pf_ns,pf_sns}).

In this article we focus on the case when the observations have low or degenerate noise.  For instance, one could consider the case:
\begin{equation}
\label{eq:add_noise_ex}
Y_n = A_n X_n + \Delta^{1/2}\epsilon_n,
\end{equation}
where $(Y_n,X_n)\in\mathbb{R}^{d_y}\times\mathbb{R}^{d_x}$,  $\Delta\geq 0$, 
$\{\epsilon_n\}_{n\geq 1}$ is a sequence of $\mathbb{R}^{d_y}$-valued independent and identically distributed (i.i.d.) noises and $A_n$ a sequence of $d_y\times d_x$ real matrices.
For technical reasons that will become apparent in the sequel, we assume that $1\leq d_y<d_x$ and that $A_n$ is of full row rank for every $n\ge 1$.
The low noise case could be interpreted as $\Delta$ small (e.g.~$\Delta=10^{-6}$) whereas the  degenerate case corresponds to $\Delta=0$.  As noticed by \cite{zhu},  such a setting is
important in many applications  where it is expected that the signal dynamics is strongly driving the observed data; see e.g.~\cite{app1,app2}.  From a mathematical perspective,  in the case $\Delta=0$ one has to define the filter on the following sequence of \emph{manifolds}:
\begin{equation*}
\mathsf{M}^\star_n=\{x\in\mathbb{R}^{d_x}:y_n-A_nx=0\}.
\end{equation*}
 Indeed this setting has been explored,  for instance,  in \cite{zhu}.  In a non-sequential context when one targets a single distribution (e.g., the smoothing law of $X_0,\dots,X_n|y_1,\dots,y_n$ with $n$ fixed), sampling on manifolds has been quite well investigated using Markov chain Monte Carlo (MCMC) in several interesting contexts; see for example
\cite{bhar,byrne,graham_st,graham,zappa}. The extension to the filtering problem seems only to be considered in \cite{zhang,zhu},  although \cite{zhang} do not consider a manifold that has been defined by the data such as $\mathsf{M}_n$ written above.  The approach of \cite{zhu} is to leverage sequential MCMC and as such it relies on the aforementioned MCMC literature.  Such methodology is particularly relevant in high-dimensions ($d_x$ large) but can be computationally expensive for moderate dimensional problems. Instead, within this latter class of filtering problems it is natural to consider PFs and this is the direction we follow in the present work.  More comparison is made in Section \ref{sec:conc}.

 In a more general context, understanding the filtering problem in this low noise regime has been the focus of many studies investigating continuous time dynamics and devising robust (extended) Kalman-Bucy type approximations of the filter \cite{brigo,katzur, milheiro, picard1, picard2} or establishing large deviations results for low $\Delta$  \cite{hijab,pardoux}, see also \cite{kutoyants} for a recent review. In general, we need to stress that the filtering distributions are typically analytically intractable except in fairly simple models,  such as ones involving linear and Gaussian dynamics.  Using approximations such as the extended Kalman filter or variants are biased and most often do not perform well for many models. Given also the increasing accuracy of many sensors used in practice it is important to design PFs that perform well for the degenerate or low observation noise case.

The application of {plain} PFs for low noise problems seldom works well due to the informative likelihoods resulting in very low weights for most particles. This is well-known and illustrated later in this article. The main idea of our work is to {reparameterize} and cast the given filtering problem as one of sampling on a sequence of manifolds, \emph{both in the degenerate setting and in the low noise one}.  
{For instance},  in the context of \eqref{eq:add_noise_ex}, for $\Delta > 0$, we work with the sequence of manifolds: 
\begin{equation*}
\widetilde{\mathsf{M}}_n=\{(x,\epsilon)\in\mathbb{R}^{d_x}\times\mathbb{R}^{d_y}:y_n-A_nx-\Delta^{1/2}\epsilon=0\}.
\end{equation*} 
A key idea then is that one can carefully define PFs on $\mathsf{M}_n$ and $\widetilde{\mathsf{M}}_n$, so that the algorithm in the low noise case will be robust by admitting a cost $\mathcal{O}(1)$  w.r.t.~$\Delta$, and will thus converge, in a proper sense, to the PF for the degenerate case of $\Delta=0$. 

\textcolor{blue}{
The degenerate case is not inherently ``more difficult'' in general, but the standard methods such as bootstrap particle filter are completely hopeless there: as $\Delta \to 0$, the likelihood $h_n(y_n|x_n)$ becomes an indicator function (or a delta measure in the limit), so satisfying the exact constraint $y_n = A_n x_n$ is a rare event under the prior. Consequently, the weights collapse to zero almost surely. The manifold reparameterisation makes the degenerate problem tractable by projecting the dynamics onto the constraint surface, allowing us to sample the free coordinates directly. Our key point is that \emph{if} one can design a good PF for the degenerate case, then the low-noise PF inherits its properties and converges to it as $\Delta \to 0$.
}

 In this paper we explore methodologies with the above properties in the presence of linear observation models such as 
\eqref{eq:add_noise_ex}. Our contributions can be summarized as follows:
\begin{itemize}
\item[(i)]We derive new PFs for cases of low and degenerate noise, within the setting of linear observation models.  We place minimal assumptions on the hidden process $\{X_n\}_{n\geq 0}$.
\item[(ii)]In the scenario of (i) and under assumptions, we prove that the $\mathbb{L}_r$-error (for a positive integer $r\ge 1$) of the PF in the low noise case will converge, as $\Delta\downarrow 0$, to the $\mathbb{L}_r$-error of the PF in the degenerate noise case.  In a less formal sense,  as the noise disappears the properties of the low noise algorithm `approach' those of the degenerate noise algorithm.  So if one seeks `good' algorithms in the low noise case,  one should start by designing a good algorithm in the  degenerate noise case.
\item[(iii)] For additive Gaussian errors both in the hidden chain and the observations, we derive the optimal proposal and we investigate 
the corresponding PFs both under low and degenerate noise case. In particular, we show that the weights of the PF for $\Delta>0$ converge  to the ones for $\Delta=0$ as $\Delta\downarrow 0$.
\item[(iv)]In the case where the hidden dynamics follow a {continuous time} diffusion process
which is observed at discrete times,  
\textcolor{blue}{we derive a new PF that is robust to low noise and employs a bridge construction, which is known to be robust to fine time discretization.}
\end{itemize}
\textcolor{blue}{Contribution (i) is delivered by Algorithms~\ref{alg:low_noise_pf} and~\ref{alg:deg_pf} in Section~\ref{sec:generic_pf}; (ii) by Proposition~\ref{res:2}; (iii) by Proposition~\ref{prop:prop}; and (iv) by Section~\ref{sec:diff}.}

The significance of this work is not only restricted to filtering.  The new PFs in this article can be used for parameter estimation
associated to diffusion models,  such as was done in \cite{diff_bridge,non_synch} using the multilevel Monte Carlo method.  We note in terms of the fourth contribution,  that \cite{bierkens} develop a methodology for the case of hypoelliptic diffusions and degenerate noise (with linear constraints),  stating that PFs can be used in their context.  However, they do not make the point that this enables the low noise case,  which is one of the main findings of this work.

This article is structured as follows.  In Section \ref{sec:model} we present the collection of models we will consider.
In Section \ref{sec:case_study} we give two particular types of models with PF algorithms and mathematical results.
In Section \ref{sec:diff} we show how to extend our methodology to the case where the hidden dynamics follow a diffusion process.
In Section \ref{sec:numerics} we present our numerical results.  
The article is concluded in Section \ref{sec:conc}.
The proofs of our mathematical results can be found in Appendix \ref{app:proof}.

\section{Modeling and Filtering}\label{sec:model}

\subsection{State-Space Model}


\textcolor{blue}{
In this section, we work in a general setting in terms of the class of models we consider. Our objective is to present what we call the \emph{noise parameterization} (as an alternative to the \emph{natural parameterization}) and the degenerate noise case in full generality. This general formulation is introduced only to define these parameterizations and to clarify the role of the observation noise. The new particle filters and theoretical results developed in this paper concern the case where the observation equation is linear in the state and has additive noise; see Section~\ref{sec:case_study}. The hidden dynamics are not assumed to be linear. Thus, the general formulation below is motivational, and the concrete observation model considered in the rest of the paper is specified in Section~\ref{sec:case_study}.
}

We consider the following state-space model (SSM) for $n\in\mathbb{N}$ and $X_0\in\mathbb{R}^{d_x}$:
\begin{align*}
Y_n & = H_n(X_n,\epsilon_n), \\
X_n & =  F_n(X_{n-1},\nu_n),
\end{align*}
where we have that: $(Y_n,X_n)\in\mathbb{R}^{d_y+d_x}$, $(d_x,d_y)\in\mathbb{N}^2$;  $\epsilon_n\in\mathbb{R}^{d_y}$, $\nu_n\in\mathbb{R}^{d_x}$ form  sequences of noises that are  independent both across $n$ and of each other, with all variables having  positive and smooth Lebesgue densities;  $H_n:\mathbb{R}^{d_x+d_y}\rightarrow\mathbb{R}^{d_y}$, $F_n:\mathbb{R}^{2d_x}\rightarrow\mathbb{R}^{d_x}$ are  sequences of  continuously differentiable functions with Jacobians that are of full row rank.

\subsection{Natural parameterization}

In the given scenario we can define the filtering densities w.r.t.~the $d_x$-dimensional Lebesgue measure as follows.
Denoting the positive conditional Lebesgue densities of $Y_n|X_n$ and $X_n|X_{n-1}$ as $h_n(y_n|x_n)$ and $f_n(x_n|x_{n-1})$ respectively, we have that the filter (see e.g.~\cite{bain} and the references therein) at time $n=1$ writes as follows (with some abuse of notation, here and elsewhere below we use the same symbol both for a distribution and for its density w.r.t.~a reference measure implied by the context): 
\begin{align}
\label{eq:f-1}
\pi_1(x_1)\equiv \frac{d\pi_1}{d\lambda_{d_x}}(x_1) =  \frac{h_1(y_1|x_1)f_1(x_1|x_0)}{\int_{\mathbb{R}^{d_x}}h_1(y_1|x_1)f_1(x_1|x_0)\lambda_{d_x}(dx_1)},
\end{align}
where $\lambda_{d_x}$ is the $d_x$-dimensional Lebesgue measure and we suppress the dependence of $y_1$ in the expression for the filtering density $\pi_1(x_1)$.
The filter at any subsequent time point $n\geq 2$ has the following prediction-updating formula:
\begin{align}
\label{eq:f-n}
\pi_n(x_n) \equiv   \frac{d\pi_n}{d\lambda_{d_x}}(x_n) =  \frac{h_n(y_n|x_n)\int_{\mathbb{R}^{d_x}} f_n(x_n|x_{n-1})\pi_{n-1}(x_{n-1})\lambda_{d_x}(dx_{n-1})}{\int_{\mathbb{R}^{d_x}}h_n(y_n|x_n)\int_{\mathbb{R}^{d_x}} f_n(x_n|x_{n-1})\pi_{n-1}(x_{n-1})\lambda_{d_x}(dx_{n-1})\lambda_{d_x}(dx_n)}.
\end{align}

In practice one cannot compute the filter  and one must resort to numerical methods. In standard settings, PFs are well known to provide accurate approximations of $\pi_n(x_n)$ and this is for an arbitrary long $n$ under favourable stability assumptions for $F_n$ and non-degeneracy of $\epsilon_n$. A PF algorithm is in fact designed to target at time $n$ the smoothing density
$$
\pi_n({x}_{1:n}) \propto \prod_{k=1}^n h_k(y_k|x_k)
f_k({x}_k|x_{k-1}),
$$
from which $\pi_n(x_n)$ can be derived as its final marginal. It should be clear that in this formulation one cannot deal directly with a degenerate case for the conditional likelihood $h_k$ and we will redefine an equivalent problem on an appropriate space. In addition, for the low noise case one should note that the region where $h_k$ has substantial value will be quite small and becomes a low probability event for traditional sampling proposals to reach it and hence standard PFs can be ineffective.
We proceed below with a different parameterization for the same SSM model which will allow for the development of effective numerical methods in the scenario of low noise for the observations.  


\subsubsection{Particle Filter for the Natural Parameterization}

We will present a generic PF for this parameterization when $\Delta>0$.  We will use the ancestor representation of the resampling indices denoted as $({a}_n^1,\dots,{a}_n^N)\in\{1,\dots,N\}^N$ and we use the convention that ${a}_0^i=i$ for each $i\in\{1,\dots,N\}$.  
For the importance sampling part, $q_n$ is a sequence of positive Markov transition densities on $\mathbb{R}^{d_x}$ and for the weights we set for any $(x',x)\in\mathbb{R}^{2d_x}$. 
$$
w_n(x',x) = \frac{h_n(y_n|x)f_n(x|x')}{q_n(x|x')}.
$$
Note that ${q}_n$ can depend on $\Delta$ and when $n=1$ we may (or may not) condition on $x_0$
in ${q}_1$ but write ${q}_1(\cdot|x_0)$ and use $x_0^i=x_0$ for all $i$.

The optimal proposal in terms of minimizing the variance of the weights conditional on ${x}_{n-1}$ is 
$$
q_n^{\text{opt}}(x_n|x_{n-1}) = \frac{h_n(y_n|x_n)f_n(x_n|x_{n-1})}
{\int_{\mathbb{R}^{d_x}} h_n(y_n|x_n)f_n(x_n|x_{n-1})\lambda_{d_x}(dx_n)}.
$$
Often used (and maligned) is the \emph{bootstrap PF} where one uses $q_n(x|x')=f_n(x|x')$.

The procedural steps of the particle filter follow the generic sequential importance resampling (SIR) structure given in Algorithm~\ref{alg:sir_core} in Section~\ref{sec:generic_pf}; the natural parameterization variants (Bootstrap and Optimal) correspond to the choices of $(q_n, w_n, Z_n)$ specified in Algorithms~\ref{alg:bootstrap_pf} and~\ref{alg:optimal_pf}.

\subsection{Noise parameterization and Degenerate Noise}

\textcolor{blue}{
The derivations below reformulate the filtering problem by treating the observation as a constraint rather than a likelihood. This remains well-defined as the noise vanishes, places the low-noise and degenerate cases in a common setting, and enables the particle filters in Section~\ref{sec:case_study} and the comparison between the two cases.
}

\subsubsection{Noise parameterization}

We set $c_n(x_n,\epsilon_n) = y_n-H_n(x_n,\epsilon_n)$ and define the manifold:
\begin{equation}
\label{eq:M-general-low}
 \widetilde{\mathsf{M}}_n=\{(x_n,\epsilon_n)\in\mathbb{R}^{d_x}\times\mathbb{R}^{d_y}:
c_n(x_n,\epsilon_n) = 0\} .
\end{equation}
Conditionally on $x_{n-1}$ (but not on $y_n$), the variables $(x_n,\epsilon_n)$ have a Lebesgue density: 
\begin{equation*}
p_n(x_{n}, \epsilon_n)  = f_n(x_n|x_{n-1})\,p_n(\epsilon_n) .
\end{equation*}
Here $p_n(\epsilon_n)$ is the Lebesgue density of $\epsilon_n$.
It is useful to think  of $p_n(x_n,\epsilon_n)$ as the prior density for $(x_n, \epsilon_n)$ given $x_{n-1}$ before consideration of $y_n$.  Once $y_n$ is received, $(x_n,\epsilon_n)$ will lie on the manifold  $\widetilde{\mathsf{M}}_n$ in (\ref{eq:M-general-low}). In particular, the posterior distribution of $(x_n,\epsilon_n)$ is their prior with its support now restricted on  $\widetilde{\mathsf{M}}_n$ (up to a constant and Jacobian factor). One can obtain an expression for the density of the posterior upon consideration of the appropriate reference measure on the manifold.  

In detail, we assume that the ambient space $\mathbb{R}^{d_x+d_y}$ is equipped with a metric tensor with a fixed positive definite matrix representation $M$. Then, it is simple to show that the filter density 
of $(x_n,\epsilon_n)$ w.r.t.~the Riemannian measure on $\widetilde{\mathsf{M}}_n$ under $M$, denoted $\mu_{\widetilde{\mathsf{M}}_n}$,
is as follows (see e.g.~\cite{zhu}).  We set:
\begin{align}
\label{eq:Gram-e}
\widetilde{G}_n(x_n,\epsilon_n) = \textrm{det}(\partial c_n(x_n,\epsilon_n)M^{-1}\partial c_n(x_n,\epsilon_n)^{\top})^{-1/2},
\end{align}
where $\partial c_n$ is the Jacobian matrix and $\textrm{det}(\cdot)$ is the determinant.
For time $n=1$ we have the filtering density:
\begin{align*}
\widetilde{\pi}_1(x_1,\epsilon_1) \equiv \frac{d\widetilde{\pi}_1}{d\mu_{\widetilde{\mathsf{M}}_1}}(x_1,\epsilon_1)  =   \frac{
\widetilde{G}_1(x_1,\epsilon_1)f_1(x_1|x_0)p_1(\epsilon_1)}{
\int_{\widetilde{\mathsf{M}}_1}
\widetilde{G}_1(x_1,\epsilon_1)f_1(x_1|x_0)p_1(\epsilon_1)
\mu_{\widetilde{\mathsf{M}}_1}(d(x_n,\epsilon_n))
}.
\end{align*}
At any subsequent time $n\geq 2$, we have:
\begin{align*}
&\widetilde{\pi}_n(x_n,\epsilon_n) \equiv \frac{d\widetilde{\pi}_{n-1}}{d\mu_{\widetilde{\mathsf{M}}_n}}(x_n,\epsilon_n)    \\ 
 &\qquad = \frac{
\widetilde{G}_n(x_n,\epsilon_n)p_n(\epsilon_n) \int_{\widetilde{\mathsf{M}}_{n-1} } f_n(x_n|x_{n-1})
\widetilde{\pi}_n(x_{n-1},\epsilon_{n-1}) \mu_{\widetilde{\mathsf{M}}_{n-1}}(d(x_{n-1},\epsilon_{n-1}))
}{
\int_{\widetilde{\mathsf{M}}_n}
\widetilde{G}_n(x_n,\epsilon_n)p_n(\epsilon_n) \int_{\widetilde{\mathsf{M}}_{n-1} } f_n(x_n|x_{n-1})
\widetilde{\pi}_{n-1}(x_{n-1},\epsilon_{n-1}) \mu_{\widetilde{\mathsf{M}}_{n-1}}(d(x_{n-1},\epsilon_{n-1}))
\mu_{\widetilde{\mathsf{M}}_n}(d(x_n,\epsilon_n))
}.
\end{align*}
We note that under the law $d\widetilde{\pi}_n(x_n, \epsilon_n)$, $n\in \mathbb{N}$,  the marginal  of $x_n$ is as given in  expressions (\ref{eq:f-1})-(\ref{eq:f-n}) for the natural parameterization. As before one can also write a path smoothing density
\begin{align}
\widetilde{\pi}_n(x_{0:n},\epsilon_{0:n}) \equiv \frac{d\widetilde{\pi}_n}{d\mu_{\bigotimes_{k=1}^n \widetilde{\mathsf{M}}_k}}(x_{0:n},\epsilon_{0:n})    \propto 
\prod_{k=1}^n \widetilde{G}_k(x_k,\epsilon_k)p_k(\epsilon_k)  f_k(x_k|x_{k-1}),
\label{eq:smoothing_M_n_low_noise}
\end{align}
which will form the target density of a PF.

We remark that one can obtain some corresponding reparameterized expressions by working  with the noise variables $(\nu_n)$ of the hidden state instead of $(x_n)$. However, this is not useful in our setting  as we focus on observational noise.  As we will establish in the next section, when the noise in the observation equation is low,  one can consider numerical algorithms under the noise parameterization and ultimately,  as the size of the noise goes to $0$,  recover an algorithm associated to the degenerate noise filtering problem.  For this latter problem, we obtain in the subsection that follows expressions analogous to the ones in the current subsection.



\subsubsection{Degenerate Noise}

We now consider degenerate observation noise and show the steps for obtaining expression for the sequence of filtering densities w.r.t.~appropriate reference measures.  We write the SSM under consideration as:
\begin{align*}
Y_n & = H_n^{\star}(X_n), \\
X_n & =  F_n(X_{n-1},\nu_n),
\end{align*}
where: $(Y_n,X_n)\in\mathbb{R}^{d_y+d_x}$, $(d_x,d_y)\in\mathbb{N}^2$, $d_y<d_x$;  $\nu_n\in\mathbb{R}^{d_x}$ is a sequence of independent noises (across $n$) with positive and smooth Lebesgue density; and $H_n^{\star}:\mathbb{R}^{d_x}\rightarrow\mathbb{R}^{d_y}$, $F_n:\mathbb{R}^{2d_x}\rightarrow\mathbb{R}^{d_x}$ are sequences of  continuously differentiable functions whose Jacobian is of full row rank.  Note that we require $d_y<d_x$ as we aim to work with an embedded manifold.

As above, we set $c_n(x_n) = y_n-H_n^{\star}(x_n)$ and define the manifold:
\begin{align}
\label{eq:M-general-zero}
 \mathsf{M}_n^{\star}=\{x_n\in\mathbb{R}^{d_x}:
c_n(x_n) = 0\}.
\end{align}
We assume that $\mathbb{R}^{d_x}$ is equipped with a metric tensor with a fixed positive definite matrix representation, denoted $M$.  We set:
\begin{equation}
\label{eq:Gram}
G_n(x_n) = \textrm{det}(\partial c_n(x_n)M^{-1}\partial c_n(x_n)^{\top})^{-1/2}.
\end{equation}
As before the filter densities w.r.t.~$\mu_{\mathsf{M}_n^{\star}}$ can be written down as
\begin{align*}
\pi_1^{\star}(x_1) \equiv 
\frac{d\pi_1^{\star}}{d\mu_{\mathsf{M}_{1}^{\star}}}(x_1) = \frac{
G_1(x_1)f_1(x_1|x_0)}{
\int_{\mathsf{M}_1^{\star}}
G_1(x_1)f_1(x_1|x_0)
\mu_{\mathsf{M}_1^{\star}}(dx_1)
}.
\end{align*}
At any subsequent time $n\geq 2$, we have: 
\begin{align*}
\pi_n^{\star}(x_n) = \frac{d\pi_n^{\star}}{d\mu_{\mathsf{M}_{n}^{\star}}}(x_n) =   \frac{
G_n(x_n) \int_{\mathsf{M}_{n-1}^{\star} } f_{n}(x_n|x_{n-1})
\pi_{n-1}^{\star}(x_{n-1}) \mu_{\mathsf{M}_{n-1}^{\star}}(dx_{n-1})
}{
\int_{\mathsf{M}_n^{\star}}
G_n(x_n)\int_{\mathsf{M}_{n-1}^{\star} } f_n(x_n|x_{n-1})
\pi_{n-1}^{\star}(x_{n-1}) \mu_{\mathsf{M}_{n-1}^{\star}}(dx_{n-1})
\mu_{\mathsf{M}_n^{\star}}(dx_n)
}.
\end{align*}
Similarly for smoothing density we have
\begin{align}
{\pi}_n^{\star}(x_{0:n}) \equiv \frac{d\pi_n^{\star}}{d\mu_{\bigotimes_{k=1}^n \mathsf{M}^{\star}_k}}(x_{0:n})    \propto 
\prod_{k=1}^n {G}_k(x_k) f_k(x_k|x_{k-1}).
\label{eq:smoothing_M_n_degenerate}
\end{align}

\section{Filtering with Linear Observation and  Additive Noise}\label{sec:case_study}

\textcolor{blue}{
In this section we consider the case where the observation equation is linear in the state with additive noise, while the hidden transition density $f_n$ is left general. We first treat general additive noise (Section \ref{sec:noise_add}) and derive PFs in the low and degenerate noise cases. We then examine (Section \ref{sec:spec_add}) the case where the hidden process also involves additive Gaussian noise, similarly to the observation process, and derive in such a setting the optimal proposals for the low and degenerate noise models.
}

\subsection{Reparameterizations and Particle Filters}\label{sec:noise_add}

We consider the SSM given as follows,  for $n\in\mathbb{N}$ and $X_0\in\mathbb{R}^{d_x}$:
\begin{align}
\label{eq:obs-linear}
Y_n  = A_n X_n + \Delta^{1/2}\epsilon_n ,
\end{align}
where  $Y_n\in\mathbb{R}^{d_y}$,  $1\leq d_y < d_x$,  $A_n\in\mathbb{R}^{d_y\times d_x}$ is of full row rank,  $\epsilon_n\in\mathbb{R}^{d_y}$ forms a sequence of independent noises with positive Lebesgue density $p_n(\cdot)$
and $\Delta\geq 0$.  For the hidden dynamics we do not impose any strong constraints beyond the existence of a  positive Lebesgue transition density which can be evaluated pointwise; it is written as $f_n(x_n|x_{n-1})$ as before. 
\begin{rem}
In this article we present noise degeneracy as the case that in the observation equation,  as $\Delta\downarrow 0$ there is no noise at all.  However,  one could consider,  say $Y_n\in\mathbb{R}^3$,  and an observation equation of the type
$$
Y_n = A_n X_n + \mathtt{P}(\Delta)\epsilon_n,
$$
where we define
$$
\mathtt{P}(\Delta)=
\begin{bmatrix}
1 & 0 & 0\\
1 & 1 & 0 \\
\Delta & \Delta & \Delta
\end{bmatrix}.  
$$
In this case one could send $\Delta\downarrow 0$ which corresponds to removing the observation noise in the $3$rd observation component. 
\textcolor{blue}{
The purpose of this remark is to note that the framework is not restricted to the case where all observation components become noiseless: this is a case of partial degeneracy, and it can be handled by the same construction with only notational changes. For ease of exposition, we therefore keep the fully degenerate scenario that we have started with.
}
\end{rem}

Recall our notation for the manifolds: 
\begin{align*}
 \widetilde{\mathsf{M}}_n&=\{(x_n,\epsilon_n)\in\mathbb{R}^{d_x+d_y}:
y_n - A_n x_n - \Delta^{1/2}\epsilon_n = 0\}; \\ 
 \mathsf{M}_n^{\star}  &=\{x_n\in\mathbb{R}^{d_x}:
y_n-A_n x_n = 0  \}.
\end{align*}
The filtering problems for the low and degenerate noise case will require sampling from  $\widetilde{\mathsf{M}}_n$ and $ \mathsf{M}_n^{\star}$ respectively. We will connect the two problems by presenting how one can view $\widetilde{\mathsf{M}}_n$  as an appropriate lifting of $ \mathsf{M}_n^{\star}$. We then discuss how this leads to generic sampling on these manifolds and then implementable PFs.

To make the geometric idea concrete before introducing the general matrix notation, consider the simple case $d_x=2$, $d_y=1$, and $A_n=(1,0)$. Writing the state as $x_n=(x_{n,1}, x_{n,2})^\top$ and the observation noise as $\epsilon_n$, the observation equation is
\begin{equation*}
y_n = x_{n,1} + \Delta^{1/2}\epsilon_n.
\end{equation*}
When $\Delta=0$, the state must satisfy $x_{n,1}=y_n$, so the degenerate manifold is simply the vertical line $x_{n,1}=y_n$, parametrized by the free coordinate $x_{n,2}=z_n\in\mathbb{R}$. For $\Delta>0$, the lifted manifold is the two-dimensional plane in $(x_{n,1},x_{n,2},\epsilon_n)$-space defined by $x_{n,1} = y_n - \Delta^{1/2}\epsilon_n$. A natural parametrization is
\begin{equation*}
x_{n,1} = y_n - \Delta^{1/2}\epsilon_n, \qquad x_{n,2} = z_n,
\end{equation*}
with $(z_n,\epsilon_n)\in\mathbb{R}^2$. As $\Delta\downarrow 0$, we recover the degenerate parametrization $(x_{n,1},x_{n,2})=(y_n,z_n)$. The general construction below gives the systematic way to perform this lifting for arbitrary dimensions $d_x$ and $d_y$.

We start from the degenerate case and let $x_n^{\star}\in \mathsf{M}_n^{\star}$ and let $V_n$ denote a $d_x\times(d_x-d_y)$ matrix whose columns form an orthonormal basis for the linear subspace $ \text{ker}(A_n)=\{x_n\in\mathbb{R}^{d_x}:A_nx_n=0\}$.
Thus, we have that:
\begin{align*}
 \mathsf{M}_n^{\star} = x_n^{\star}  + \mathscr{L}V_n,
\end{align*}
with $\mathscr{L}$ denoting linear span. The point to note here is that sampling on $\mathsf{M}_n^{\star}$ can performed by centering on  any $x_n^{\star}\in\mathsf{M}_n^{\star}$ and adding a random sample defined on $ \mathscr{L}V_n$. Due to the linear constraint from the observation, our state lives in a lower dimensional manifold, so one can proceed by appropriately sampling $z_n\in \mathbb{R}^{d_x-d_y}$ and then return the linear map $u(z_n)=x_n^{\star}  +V_n z_n$ as  a sample in $\mathsf{M}_n^{\star}$. 
\textcolor{blue}{
The standard, numerically stable way to compute $V_n$ is using the singular value decomposition (SVD) or the QR decomposition; see \cite{golub}, Section~5.4.2 (The Null Space) or Section~2.6 (The Singular Value Decomposition). These references detail how the right singular vectors corresponding to zero (or near-zero) singular values form an orthonormal basis for the kernel.
}

We `uplift' the manifold $\mathsf{M}_n^{\star}$ to obtain  $\widetilde{\mathsf{M}}_n$ by working as follows. Consider starting from the case without noise present ($\varepsilon_n=0$) and augment $x_n^{\star}$ by setting  $(\widetilde{x}_n^{\Delta})^{\top}=((x_n^{\star})^{\top},0^{\top})$, which is a $(d_x+d_y)-$dimensional vector that obeys the defining constraint of $\widetilde{\mathsf{M}}_n$, i.e. $y_n=[A_n,\Delta^{1/2}\mathtt{I}_{d_y}]\widetilde{x}_n^{\Delta}$. 
Let $V_n^{\Delta}$ be a $(d_x+d_y)\times d_x$ matrix whose columns are a basis of $\text{ker}([A_n,\Delta^{1/2}\mathtt{I}_{d_y}])$. We will set
\textcolor{blue}{
\begin{align}
V_n^\Delta =
\begin{pmatrix}
V_n & -\Delta^{1/2} A_n^\top (A_n A_n^\top)^{-1} \\
0 & I_{d_y}
\end{pmatrix}.
\label{eq:V_n_Delta}
\end{align}
With this choice, the corresponding noise coordinate is exactly the observation noise: $u_n^\Delta(\widetilde z_n,\epsilon)=\overline{z}_n$.
}
Although $V_n^{\Delta}$ is not orthonormal in general, the induced Gram matrix is constant in the coordinates and therefore cancels in the normalized importance weights and target densities. We then can write 
\begin{align*}
\widetilde{\mathsf{M}}_n = \widetilde{x}_n^{\Delta}  + \mathscr{L}\big\{V_n^{\Delta}\big\}.
\end{align*}

\hspace{-0.15cm}
We refer the reader for more details to the Appendix. From a practical point of view at each time $n$ one needs to sample a signal and an observation component that lie on the manifold  $\widetilde{\mathsf{M}}_n$. The first step is to compute a point $x_n^{\star}$ solving $A_n x_n^{\star,0}=y_n$ and thus obtain  $\widetilde{x}_n^{\Delta}$ by augmenting with $0$. Then one should sample appropriately $\widetilde{z}\in\mathbb{R}^{d_x}$ and use the following mapping: 
\begin{align}
u_n^{\Delta}(\widetilde{z}) = \widetilde{x}_n^{\Delta} + V_n^{\Delta} \widetilde{z},\label{eq:u_Delta}
\end{align}
where we define  $u_n^{\Delta}(\widetilde{z})^{\top} = (u_n^{\Delta}(\widetilde{z},x)^{\top},u_n^{\Delta}(\widetilde{z},\epsilon)^{\top})$, with
$u_n^{\Delta}(\widetilde{z},x)\in\mathbb{R}^{d_x}$ and $u_n^{\Delta}(\widetilde{z},\epsilon)\in\mathbb{R}^{d_y}$, so that in the notation $u_n^{\Delta}(\widetilde{z},x)$ is meant to indicate the first $d_x$ dimensions of $u_n^{\Delta}(\widetilde{z})$ and  $u_n^{\Delta}(\widetilde{z},\epsilon)$ the last $d_y$ dimensions,  corresponding to signal and observation noise respectively.

In both the low noise and degenerate filtering we are primarily interested for the state variables and given this discussion it should be clear one can equivalently define the low noise filtering problem w.r.t  a random variable $\widetilde{Z}_n\in\mathbb{R}^{d_x}$.  In order to better connect the two problems  we will also use throughout the following notation convention:  $\widetilde{z}_n^{\top} = (z_n^{\top},\overline{z}_n^{\top})^{\top}$,  with $(z_n,\overline{z}_n)\in \mathbb{R}^{d_x-d_y}\times \mathbb{R}^{d_y}$, recalling that $z_n$ is the random variable of interest for the degenerate case.

\subsubsection{The Low Noise Case}\label{sec:ln_add}

PFs are defined to target a sequence of smoothing distributions, so using \eqref{eq:u_Delta} we will rewrite \eqref{eq:smoothing_M_n_low_noise} as a function of $\widetilde{z}_{1:n}$ . Note that due to the manifolds considered in this section being linear, the Gram matrices in (\ref{eq:Gram-e}), (\ref{eq:Gram}) are constant and we can write densities w.r.t.~Lebesgue measure.   Given this construction,  at any time $n\geq 1$ we can write the joint smoothing density of $\widetilde{z}_{1:n}\in\mathbb{R}^{nd_x}$ w.r.t.~$\bigotimes_{i=1}^n \lambda_{d_x}(d\widetilde{z}_i)$ as 
$$
\pi_n^{\Delta}(\widetilde{z}_{1:n}) \propto \prod_{k=1}^n p_k(u_k^{\Delta}(\widetilde{z}_k,\epsilon))
f_k(u_k^{\Delta}(\widetilde{z}_k,x)|u_{k-1}^{\Delta}(\widetilde{z}_{k-1},x))
$$
where we set $u_{0}^{\Delta}(z,x)=x_0$, for any $z\in\mathbb{R}^{d_x}$.  We are assuming the probability density is well-defined for any $\Delta> 0$, that is,  the normalization is finite.

In such a scenario,  one can use a standard particle filter,  which we now detail.  Let $\widetilde{q}_n^{\Delta}$ be a sequence of positive Markov transition densities on $\mathbb{R}^{d_x}$ and set for any $(\widetilde{z}',\widetilde{z})\in\mathbb{R}^{2d_x}$
\begin{equation}
w_n^{\Delta}(\widetilde{z}',\widetilde{z}) = 
\frac{p_n(u_n^{\Delta}(\widetilde{z},\epsilon))
f_n(u_n^{\Delta}(\widetilde{z},x)|u_{n-1}^{\Delta}(\widetilde{z}',x))}
{\widetilde{q}_n^{\Delta}(\widetilde{z}\mid \widetilde{z}')}.
\label{eq:low_weight}
\end{equation}
As for the natural parameterization case, note that $\widetilde{q}_n^{\Delta}$ can depend on $\Delta$ and when $n=1$ we may (or may not) condition on $x_0$
in $\widetilde{q}_1^{\Delta}$ but write $\widetilde{q}_1^{\Delta}(\cdot|x_0)$ from herein.
Let $N\in\mathbb{N}$ be given and denote for any $n\geq 1$,  $\widetilde{z}_n^{1:N}=(\widetilde{z}_n^{1}\dots,\widetilde{z}_n^{N})$ which represent the particles of the PF.   It will prove useful to use the ancestor representation of the resampling indices so we will also be interested in $(\widetilde{a}_n^1,\dots,\widetilde{a}_n^N)\in\{1,\dots,N\}^N$ and again we use the convention that $\widetilde{a}_0^i=i$ for each $i\in\{1,\dots,N\}$.  We will also write $\widetilde{z}_{0}^{\widetilde{a}_{0}^j}=x_0$.  


Let $\varphi:\mathbb{R}^{d_x}\rightarrow\mathbb{R}$ be a bounded and measurable function (write such functions as $\mathtt{B}_b(\mathbb{R}^{d_x})$) then we can estimate:
$$
\pi_n^{\Delta}(\varphi) := \int_{\mathbb{R}^{nd_x}}\varphi(\widetilde{z}_n) \pi_n^{\Delta}(\widetilde{z}_{1:n})
\bigotimes_{i=1}^n \lambda_{d_x}(d\widetilde{z}_i)
$$
by using the estimator obtained just before the end of the iterate step
\begin{equation}
\pi_n^{\Delta,N}(\varphi) = 
\frac{\sum_{i=1}^N \varphi(\widetilde{z}_n^i)
w_{n}^{\Delta}(\widetilde{z}_{n-1}^{\widetilde{a}_{n-1}^i},\widetilde{z}_{n}^{i})}
{\sum_{i=1}^N w_{n}^{\Delta}(\widetilde{z}_{n-1}^{\widetilde{a}_{n-1}^i},\widetilde{z}_{n}^{i})}.
\label{eq:low_estimator}
\end{equation}
The convergence of this estimator and the choice of proposal have been extensively discussed in the literature; see for instance \cite{bain,delmoral} and the references therein.  We remark that the optimal,  in terms of minimizing the variance of the weights conditional on $\widetilde{z}_{n-1}$, is
$$
\widetilde{q}_n^{\Delta,\text{opt}}(\widetilde{z}_n|\widetilde{z}_{n-1}) = 
\frac{p_n(u_n^{\Delta}(\widetilde{z}_n,\epsilon))
f_n(u_n^{\Delta}(\widetilde{z}_n,x)|u_{n-1}^{\Delta}(\widetilde{z}_{n-1},x))}
{\int_{\mathbb{R}^{d_x}}
p_n(u_n^{\Delta}(\widetilde{z}_n,\epsilon))
f_n(u_n^{\Delta}(\widetilde{z}_n,x)|u_{n-1}^{\Delta}(\widetilde{z}_{n-1},x))\lambda_{d_x}(d\widetilde{z}_n)
}.
$$
This is tractable only in certain modeling scenarios, one of which we investigate below.  We remark that in the case that one
is able to compute the optimal proposal,  due to the one-to-one mappings from the natural parameterization to the noise parameterization to the map back to the Lebesgue measure in the linear constraint scenario,  the optimal proposal from the natural parameterization will provide an equivalent formulation.  One of the main points of the low noise method however,  is that it will permit its application in cases where the optimal proposal is not available and will be robust to low values of the noise parameter.

The procedural steps of the low-noise particle filter follow the generic SIR structure given in Algorithm~\ref{alg:sir_core} in Section~\ref{sec:generic_pf}, with the proposal, weight, and state coordinate specified in Algorithm~\ref{alg:low_noise_pf}.

\subsubsection{The Degenerate Noise Case}\label{sec:dn_add}

The degenerate case can be handled in a similar manner,  which we now describe.
For any $(n,z)\in\mathbb{N}\times\mathbb{R}^{d_x-d_y}$ set
$$
u_n^{\star}(z) = x_n^{\star} + V_n z
$$
with the convention that $u_0^{\star}(z)=x_0$ for any $z\in \mathbb{R}^{d_x-d_y}$.
At any time $n\geq 1$ we can write the joint smoothing density of $z_{1:n}\in\mathbb{R}^{n(d_x-d_y)}$ w.r.t.~$\bigotimes_{i=1}^n \lambda_{d_x-d_y}(dz_i)$ as 
$$
\pi_n^{\star}(z_{1:n}) \propto \prod_{k=1}^n 
f_k(u_k^{\star}(z_k)|u_{k-1}^{\star}(z_{k-1})),
$$
where we note that $u_{0}^{\star}(z_{0}) = x_0$. Let $q_n^{\star}$ be a sequence of positive Markov transition densities on $\mathbb{R}^{d_x-d_y}$ and set for any $(z',z)\in\mathbb{R}^{2(d_x-d_y)}$
\begin{equation}
w_n^{\star}(z',z) = 
\frac{f_n(u_n^{\star}(z)\mid u_{n-1}^{\star}(z'))}
{q_n^{\star}(z\mid z')}.
\label{eq:deg_weight}
\end{equation}
When $n=1$ we may (or may not) condition on (some part of) $x_0$
in $q_n^{\star}$ but again write $q_n^{\star}(\cdot|x_0)$ from herein.  Denote for any $n\geq 1$,  $z_n^{1:N}=(z_n^{1}\dots,z_n^{N})$, the particles of the PF.   The ancestor variables are $(a_n^1,\dots,a_n^N)\in\{1,\dots,N\}^N$ and we use the convention that $a_0^i=i$ for each $i\in\{1,\dots,N\}$.  We will also write $z_{0}^{a_{0}^j}=x_0$.


Let $\varphi\in\mathtt{B}_b(\mathbb{R}^{d_x-d_y})$ then we can estimate:
$$
\pi_n^{\star}(\varphi) := \int_{\mathbb{R}^{n(d_x-d_y)}}\varphi(z_n) \pi_n^{\star}(z_{1:n})
\bigotimes_{i=1}^n \lambda_{d_x-d_y}(dz_i)
$$
by using the estimator, obtained just before the end of the iterate step
\begin{equation}
\pi_n^{\star,N}(\varphi) := 
\frac{\sum_{i=1}^N \varphi(z_n^i)
w_{n}^{\star}(z_{n-1}^{a_{n-1}^i},z_{n}^{i})}
{\sum_{i=1}^N
w_{n}^{\star}(z_{n-1}^{a_{n-1}^i},z_{n}^{i})}.
\label{eq:deg_estimator}
\end{equation}
As in the low noise case,  the optimal proposal here would be
$$
q_n^{\star,\text{opt}}(z_n|z_{n-1}) = 
\frac{f_n(u_n^{\star}(z_n)|u_{n-1}^{\star}(z_{n-1}))}
{\int_{\mathbb{R}^{d_x-d_y}}
f_n(u_n^{\star}(z_n)|u_{n-1}^{\star}(z_{n-1}))\lambda_{d_x-d_y}(dz_n)
}.
$$

The procedural steps of the degenerate-noise particle filter follow the generic SIR structure given in Algorithm~\ref{alg:sir_core} in Section~\ref{sec:generic_pf}, with the proposal, weight, and state coordinate specified in Algorithm~\ref{alg:deg_pf}.

\subsubsection{A Low Noise Convergence Result}

We now consider the convergence,  in a sense to be made precise below,  of the PF in the low noise case to that of the degenerate noise case.  
Recall that we use the notation for $\widetilde{z}\in\mathbb{R}^{d_x}$ as $\widetilde{z}^{\top}=(z^{\top},\overline{z}^{\top})$, with $z\in\mathbb{R}^{d_x-d_y}$.
We will constrain $\Delta\in[0,\overline{\Delta}]$, $\overline{\Delta}>0$ and this is reasonable as we are interested in the low noise problem.
We make the following assumption.
\begin{hypA}\label{ass:1}
\begin{enumerate}
\item{For any $(n,\widetilde{z}',\widetilde{z})\in\mathbb{N}\times\mathbb{R}^{2d_x}$ we have
$$
\lim_{\Delta\downarrow 0} \widetilde{q}_n^{\Delta}(\widetilde{z}|\widetilde{z}') = 
p_n(\overline{z}) q_n^{\star}(z|z').
$$}
\item{For any $n\in\mathbb{N}$,  $p_n(\epsilon)f_n(x|x')$ is a continuous function on $\mathbb{R}^{d_y+2d_x}$.}
\item{For any $n\in\mathbb{N}$ there exists a $C<+\infty$ such that
$$
\sup_{(\Delta,\widetilde{z}',\widetilde{z})\in[0,\overline{\Delta}]\times\mathbb{R}^{2d_x}} 
\frac{\widetilde{q}_n^{\Delta}(\widetilde{z}|\widetilde{z}')}{p_n(\overline{z}) q_n^{\star}(z|z')} \leq C.
$$
}
\item{For any $n\in\mathbb{N}$ there exists a $C<+\infty$ such that
$$
\sup_{(\Delta,\widetilde{z}',\widetilde{z})\in[0,\overline{\Delta}]\times\mathbb{R}^{2d_x}} w_n^{\Delta}(\widetilde{z}',\widetilde{z}) \leq C.
$$
}
\end{enumerate}
\end{hypA}
We discuss the assumptions in turn.  1.~is not overly restrictive as one chooses the proposal densities.  The assumption means that as the noise disappears,  one should propose the $d_x-d_y$ random variable,  just as in the degenerate noise case and the noise variable according to the noise density. 2.~is needed as we will appeal to various pointwise limits and indeed could be weakened to continuity at a point; we simply use global continuity for simplicity.  3.~simply says that the limiting proposal is a good importance proposal for the low-noise case proposal,  globally in $\Delta$; by 1.~it is automatically satisfied for $\Delta=0$. 4.~states that the importance weights are upper-bounded and is generally an assumption needed for importance sampling algorithms to work well.  It can hold under well chosen proposals.

We denote by $\mathbb{E}^{\Delta}$ as expectations w.r.t.~the law of the PF in the low noise case and 
by $\mathbb{E}^{\star}$ as expectations w.r.t.~the law of the PF in the degenerate noise case.  We can now state our main result whose proof is in Appendix \ref{app:res2}.
\begin{prop}\label{res:2}
Assume (A\ref{ass:1}).  Then for any $(n,r,\varphi)\in\mathbb{N}^2\times\mathtt{B}_b(\mathbb{R}^{d_x-d_y})$ we have
$$
\lim_{\Delta \downarrow 0}\mathbb{E}^{\Delta}\left[\left|\pi_n^{\Delta,N}(\varphi)-\pi_n^{\Delta}(\varphi)\right|^r\right]^{1/r} = 
\mathbb{E}^{\star}\left[\left|\pi_n^{\star,N}(\varphi)-\pi_n^{\star}(\varphi)\right|^r\right]^{1/r}.
$$
\end{prop}

\begin{rem}\label{rem:1}
The main purpose of Proposition \ref{res:2} is to verify that the properties of the low noise case PF are inherited,  as 
$\Delta \downarrow 0$ from that of the degenerate noise case.  In other words if one can design a good PF for the 
degenerate noise case,  then the low noise filtering problem can be effectively handled. 
\textcolor{blue}{
For instance, under multiplicative drift conditions, one has some time uniform stability of the PF \cite{whiteley} and if these hold for the degenerate case, they are likely to hold as  $\Delta \downarrow 0$. 
This should not be interpreted as a claim that the degenerate problem is intrinsically easier; it is simply the limiting case in which the constraint is exact and the manifold parametrization is most transparent.
}
\end{rem}

\subsubsection{Generic PF Framework}\label{sec:generic_pf}

We now summarize the particle filters developed in this section. Algorithm~\ref{alg:sir_core} provides a generic SIR structure written in terms of the particle coordinate $Z_n$. The four variants considered in this paper - bootstrap, optimal natural, low-noise manifold, and degenerate manifold - are obtained by specifying the proposal $q_n$, incremental weight $w_n$, and particle coordinate $Z_n$ in Algorithm~\ref{alg:sir_core}, as detailed in Algorithms~\ref{alg:bootstrap_pf}-\ref{alg:deg_pf}. In the natural parameterization $Z_n=X_n$; in the low-noise manifold case $Z_n=\widetilde Z_n$; and in the degenerate manifold case $Z_n\in\mathbb{R}^{d_x-d_y}$. For the manifold-based variants, the maps $u_n^\Delta$ and $u_n^\star$ from Section~\ref{sec:noise_add}, and the incremental weights $w_n^\Delta$ and $w_n^\star$ from Sections~\ref{sec:ln_add} and~\ref{sec:dn_add}, are used.

\begin{algorithm}[h]
\caption{Generic SIR Core}
\label{alg:sir_core}
\begin{algorithmic}[1]
\Require $N$, proposal $q_n$, weight $w_n$, data $\{y_n\}$, $x_0$.
\State Set $a_0^i=i$, $i=1,\ldots,N$, and $n=1$.
\While{new observation $y_n$}
    \State \textbf{Propagate:} If $n=1$, sample
    $Z_1^i \sim q_1(\cdot \mid x_0)$, $i=1,\ldots,N$; otherwise sample
    $Z_n^i \sim q_n(\cdot \mid Z_{n-1}^{a_{n-1}^i})$, $i=1,\ldots,N$.
    \State \textbf{Weight:} If $n=1$, compute
    $w_1^i = w_1(x_0,Z_1^i)$, $i=1,\ldots,N$; otherwise compute
    $w_n^i = w_n(Z_{n-1}^{a_{n-1}^i}, Z_n^i)$, $i=1,\ldots,N$.
    \State \textbf{Normalize:}
    $\rho_n^i = \frac{w_n^i}{\sum_{s=1}^N w_n^s}$, $i=1,\ldots,N$.
    \State \textbf{Estimate:}
    $\pi_n^N(\varphi) = \sum_{i=1}^N \rho_n^i\,\varphi(Z_n^i)$.
    \State \textbf{Resample:} Sample
    $a_n^i \sim \mathrm{Cat}(\cdot; N,\{\rho_n^s\}_{s=1}^N)$, $i=1,\ldots,N$.
    \State Set $n \leftarrow n+1$.
\EndWhile
\end{algorithmic}
\end{algorithm}

\begin{algorithm}[h]
\caption{Bootstrap Particle Filter}
\label{alg:bootstrap_pf}
\begin{algorithmic}[1]
\Require $N, f_n, h_n, \{y_n\}, x_0$.
\State Use Algorithm~\ref{alg:sir_core} with $Z_n=X_n$ and, for $(x',x)\in\mathbb{R}^{2d_x}$,
$$
q_n(x\mid x') = f_n(x\mid x'),
\qquad
w_n(x',x) = h_n(y_n\mid x).
$$
\end{algorithmic}
\end{algorithm}

\begin{algorithm}[h]
\caption{Optimal Particle Filter (Natural)}
\label{alg:optimal_pf}
\begin{algorithmic}[1]
\Require $N, f_n, h_n, \{y_n\}, x_0$.
\State Use Algorithm~\ref{alg:sir_core} with $Z_n=X_n$ and, for $(x',x)\in\mathbb{R}^{2d_x}$,
$$
q_n(x\mid x') =
\frac{h_n(y_n\mid x)f_n(x\mid x')}
{\int_{\mathbb{R}^{d_x}} h_n(y_n\mid \bar x)f_n(\bar x\mid x')\,d\bar x},
$$
$$
w_n(x',x) =
\int_{\mathbb{R}^{d_x}} h_n(y_n\mid \bar x)f_n(\bar x\mid x')\,d\bar x.
$$
\end{algorithmic}
\end{algorithm}

%



\begin{algorithm}[h]
\caption{Low-Noise Manifold PF}
\label{alg:low_noise_pf}
\begin{algorithmic}[1]
\Require $N, f_n, p_n, \widetilde q_n^\Delta, \{y_n\}, x_0, u_n^\Delta$.
\State Use Algorithm~\ref{alg:sir_core} with $Z_n=\widetilde Z_n$ and, for $(\widetilde z',\widetilde z)\in\mathbb{R}^{2d_x}$,
$$
q_n(\widetilde z\mid \widetilde z') =
\widetilde q_n^\Delta(\widetilde z\mid \widetilde z'),
$$
$$
w_n(\widetilde z',\widetilde z) =
\frac{
    p_n\bigl(u_n^\Delta(\widetilde z,\epsilon)\bigr)
    f_n\bigl(u_n^\Delta(\widetilde z,x)\mid u_{n-1}^\Delta(\widetilde z',x)\bigr)
}{
    \widetilde q_n^\Delta(\widetilde z\mid \widetilde z')
}.
$$
\State The corresponding state particle is
$$
X_n = u_n^\Delta(\widetilde Z_n,x).
$$
\end{algorithmic}
\end{algorithm}

\begin{algorithm}[h]
\caption{Degenerate Manifold PF}
\label{alg:deg_pf}
\begin{algorithmic}[1]
\Require $N, f_n, q_n^\star, \{y_n\}, x_0, u_n^\star$.
\State Use Algorithm~\ref{alg:sir_core} with $Z_n\in\mathbb{R}^{d_x-d_y}$ and, for $(z',z)\in\mathbb{R}^{2(d_x-d_y)}$,
$$
q_n(z\mid z') = q_n^\star(z\mid z'),
$$
$$
w_n(z',z) =
\frac{
    f_n\bigl(u_n^\star(z)\mid u_{n-1}^\star(z')\bigr)
}{
    q_n^\star(z\mid z')
}.
$$
\State The corresponding state particle is
$$
X_n = u_n^\star(Z_n).
$$
\end{algorithmic}
\end{algorithm}



\subsection{Additive Gaussian Noise with Nonlinear Signal Dynamics}\label{sec:spec_add}

We consider the following state-space model for $n\in\mathbb{N}$ and $X_0\in\mathbb{R}^{d_x}$ given:
\begin{align*}
Y_n & = A_n X_n + \Delta^{1/2}\epsilon_n \\
X_n & =  f_n(X_{n-1}) + \nu_n,
\end{align*}
where $Y_n\in\mathbb{R}^{d_y}$,  $1\leq d_y < d_x$,  $A_n\in\mathbb{R}^{d_y\times d_x}$ is of full row rank, 
$\Delta\geq 0$
$f_n:\mathbb{R}^{d_x}\rightarrow\mathbb{R}^{d_x}$. Independently for each $n\in\mathbb{N}$ $\epsilon_n\sim\mathcal{N}_{d_y}(0,\Sigma_n)$ and independently $\nu_n\sim\mathcal{N}_{d_x}(0,\Omega_n)$ where, $\Sigma_n,\Omega_n$ are a sequence of real positive definite symmetric matrices. 
Here  $\mathcal{N}_d(\mathsf{m},\mathsf{K})$ denotes the $d-$dimensional Gaussian distribution with mean $\mathsf{m}$ and covariance matrix $\mathsf{K}$.

\subsubsection{Optimal Proposals}

We now consider the optimal or bootstrap proposals and the importance weights in the natural parameterization,
noise parameterization (low noise) and the degenerate noise case.   To aid the notations in the below statement
we partition the matrix $V_n^{\Delta}$ into its first $d_x$ rows denoted by the $d_x\times d_x$ matrix $V_n^{\Delta}(x)$ and its last $d_y$ rows denoted by the $d_y\times d_x$ matrix $V_n^{\Delta}(\epsilon)$.  We will also use the notation
$x_0^{\star}+V^{\Delta}_0(x)z=x_0$ 
$x_0^{\star}+V_0z=x_0$
and where we have not defined the objects on left hand side.
We have the following result whose proof is in Appendix \ref{app:inc}.
 
\begin{prop}\label{prop:prop}
Consider particle filters associated to the model at the start of Section \ref{sec:spec_add}.
\begin{enumerate}
\item{Natural parameterization with bootstrap proposal:
\begin{align*}
X_n | x_{n-1} & \sim \mathcal{N}_{d_x}(f_n(x_{n-1}),\Omega_n) \\
w_n(x_n) & = \exp\left\{-\frac{1}{2\Delta}(y_n-A_nx_n)^{\top}\Sigma_n^{-1}(y_n-A_nx_n)\right\}
\end{align*}
}
\item{Low noise with optimal proposal:
\begin{align*}
\widetilde{Z}_n|\widetilde{z}_{n-1} & \sim \mathcal{N}_{d_x}(\widetilde{\mu}_n,\widetilde{\Omega}_n) \\
w_n^{\Delta,\text{\emph{opt}}}(\widetilde{z}_{n-1}) & = \exp\left\{\frac{1}{2}
\left(\widetilde{\mu}_n^{\top}\widetilde{\Omega}_n^{-1}\widetilde{\mu}_n -
\widetilde{\mathtt{m}}_n^{\top}\Omega_n^{-1}\widetilde{\mathtt{m}}_n
\right)
\right\}
\end{align*}
where
\begin{align*}
\widetilde{\mu}_n & = \widetilde{\Omega}_nV_n^{\Delta}(x)^{\top}\Omega_n^{-1}(f_n(x_{n-1}^{\star}+V_{n-1}^{\Delta}(x)\widetilde{z}_{n-1})-x_n^{\star})
\\
\widetilde{\Omega}_n^{-1} & =  V_n^{\Delta}(\epsilon)^{\top}\Sigma_n^{-1} V_n^{\Delta}(\epsilon) + 
 V_n^{\Delta}(x)^{\top}\Omega_n^{-1} V_n^{\Delta}(x) \\
\widetilde{\mathtt{m}}_n & = f_n(x_{n-1}^{\star}+V_{n-1}^{\Delta}(x)\widetilde{z}_{n-1}) - x_n^{\star}.
\end{align*}
}
\item{Degenerate noise with optimal proposal:
\begin{align*}
Z_n|z_{n-1} & \sim \mathcal{N}_{d_x-d_y}(\mu_n^{\star},\Omega_n^{\star}) \\
w_n^{\star,\text{\emph{opt}}}(z_{n-1}) & = \exp\left\{\frac{1}{2}
\left((\mu_n^{\star})^{\top}(\Omega_n^{\star})^{-1}\mu_n^{\star} -
(\mathtt{m}_n^{\star})^{\top}\Omega_n^{-1}\mathtt{m}_n^{\star}
\right)
\right\}
\end{align*}
where
\begin{align*}
\mu_n^{\star} & = \Omega_n^{\star}V_n^{\top}\Omega_n^{-1}(f_n(x_{n-1}^{\star}+V_{n-1}z_{n-1})-x_n^{\star})
\\
(\Omega_n^{\star})^{-1} & =  
 V_n^{\top}\Omega_n^{-1} V_n \\
\mathtt{m}_n^{\star} & = f_n(x_{n-1}^{\star}+V_{n-1}z_{n-1}) - x_n^{\star}.
\end{align*}
}
\end{enumerate}
Finally,  we have
$$
\lim_{\Delta \downarrow 0}
w_n^{\Delta,\text{\emph{opt}}}(\widetilde{z}_{n-1}) = w_n^{\star,\text{\emph{opt}}}(z_{n-1}).
$$
\end{prop}

\begin{rem}\label{rem:rmk_lin}
The significance of Proposition \ref{prop:prop} is as follows.  $\Delta$ controls the amount of noise and the low noise case is as $\Delta\downarrow 0$.   In the case of the bootstrap PF it is clear that numerically that $w_n(x_n)$ will degenerate to zero and hence this algorithm is of little use; this is unsurprising as the proposal does not take into account any information associated to the data.  In the low and degenerate noise cases (2 \& 3 in Proposition \ref{prop:prop}) one can characterize the optimal proposal and the associated weight and, as was the case in Proposition \ref{res:2},  there is a convergence of the weights from the low noise to degenerate case with a similar conclusion to Remark \ref{rem:1}.  As we noted before the optimal proposal in the natural parameterization will be equivalent to the one in the noise parameterization. 
\end{rem}

\section{Partially Observed Diffusions}\label{sec:diff}

\subsection{Diffusion Process}

We will now consider the problem of filtering diffusion processes in the low and degenerate noise case.
We consider the following diffusion process on the filtered probability space $(\Omega,\mathscr{F},\{\mathscr{F}_t\}_{t\geq 0},\mathbb{P})$.  
\begin{equation}
dX_t = \mu(X_t)dt + \sigma(X_t)dW_t\quad\quad X_0=x_0\label{eq:diff}
\end{equation}
where $X_t\in\mathbb{R}^{d_x}$,   
$\mu:\mathbb{R}^{d_x} \rightarrow\mathbb{R}^{d_x}$, 
and $\sigma:\mathbb{R}^{d_x} \rightarrow\mathbb{R}^{d_x\times d_x}$
and $\{W_t\}_{t\geq 0}$ is a standard $d_x-$dimensional Brownian motion.
Denote $\mu^j$ as the $j$-th component of $\mu$, $j\in\{1,\dots,d_x\}$
and $\sigma^{j,k}$ as the $(j,k)$-th component of $\sigma$, $(j,k)\in\{1,\dots,d_x\}^2$.
As the exposition will closely follow the works in \cite{diff_bridge,beskos,non_synch,stanton} we make the following assumption which is also made in most of those papers.  Below $\mathcal{C}(\mathbb{R}^{d_x})$ are the twice continuously differentiable functions on $\mathbb{R}^{d_x}$ and $\|\cdot\|$ is the Euclidean norm.
\begin{quote} 
We have $\mu^j\in \mathcal{C}^2(\mathbb{R}^{d_x})$ and $\sigma^{j,k} \in \mathcal{C}^2(\mathbb{R}^{d_x})$, for $(j,k)\in\{1,\ldots, d_x\}^2$. 
In addition, $\mu$ and $\sigma$ satisfy 
\begin{itemize}
\item[(i)] {\bf uniform ellipticity}: $\Sigma(x):=\sigma(x)\sigma(x)^{\top}$ is uniformly positive definite 
over $x\in \mathbb{R}^{d_x}$;
\item[(ii)] {\bf globally Lipschitz}:
There exists a constant $C<\infty$ such that 
$|\mu^j(x)-\mu^j(x')|+|\sigma^{j,k}(x)-\sigma^{j,k}(x')| \leq C \|x-x'\|$ 
for all $(x,x') \in \mathbb{R}^{d_x}\times\mathbb{R}^{d_x}$, $(j,k)\in\{1,\dots,d_x\}^2$. 
\end{itemize}
\end{quote}
We note that one could use the Euler-Maruyama time-discretization method to obtain the transition density of the hidden process.  However,  it is well-known (e.g.~\cite{beskos} or \cite[Figure 1]{shouto}) that this method, when combined with a PF,  typically collapses (due to increasing Monte Carlo variance) as the time-discretization becomes finer and thus we consider a robust reformulation using a bridge construction that avoids this issue and is often used in the literature \cite{schauer,vd_meulen_guided_mcmc}. Both approaches will need to be discretized in time, but will opt for the latter due to its robustness to fine time discretizations. As a result the PFs that we will present in Sections \ref{sec:ln_diff} and \ref{sec:dn_diff} 
\textcolor{blue}{are robust to low observation noise and are designed to mitigate weight degeneracy associated with fine time discretizations.}

\subsection{Bridge Construction}\label{sec:bridge}

We review the method in \cite{schauer,vd_meulen_guided_mcmc} as has been used in \cite{beskos}.
The exposition closely follows that in \cite{diff_bridge,beskos,non_synch}.
Consider the case $t\in[s_1,s_2]$, $0\leq s_1<s_2\leq T$ and let
$\mathbf{X}_{[s_1,s_2]}:=\{X_{t}\}_{t\in[s_1,s_2]}$, and $\mathbf{W}_{[s_1,s_2]}:=\{W_{t}\}_{t\in[s_1,s_2]}$.  
We will use a measure disintegration of the diffusion path measure with respect to a sampled value at time $s_2$.
Suppose $f_{t,s_{2}}(x'|x)$ denotes the unknown transition density from time $t$ to $s_2$ associated to \eqref{eq:diff}.
An equivalent formulation for the law of $\mathbf{X}_{[s_1,s_2]}$ would be to sample from $f_{s_1,s_2}$ to obtain $(x, x')\in \mathbb{R}^{2d_x}$ and then to interpolate between $x$ and $x'$ or "fill in" the path using an alternative diffusion that is conditioned to 
arrive at point $x'$ at time $s_2$. 
We will refer to this diffusion process as the bridge process and its dynamics will be as in \eqref{eq:diff}, but instead its drift will be given by $\mu(x)+\Sigma(x)\nabla_x\log{f}_{t,s_2}(x'|x)$.
Let $\mathbb{P}_{x,x'}$ denote the law of the solution of the SDE \eqref{eq:diff}, on $[s_1,s_2]$, started at $x$ and conditioned to hit $x'$ at time $s_2$.

We introduce a user-specified auxiliary process $\{\tilde{X}_t\}_{t\in[s_1,s_2]}$ following:
\begin{align}
\label{eq:aux_SDE_tilde}
d \tilde X_{t} = \tilde \mu(t,\tilde X_{t})dt + \tilde \sigma(t,\tilde X_{t})dW_{t}, \quad t\in[s_1,s_2],\quad~\tilde{X}_{s_1} =x, 
\end{align}
where $\tilde \mu:[s_1,s_2]\times\mathbb{R}^{d_x}\rightarrow\mathbb{R}^{d_x}$ and $\tilde \sigma:[s_1,s_2]\times\mathbb{R}^{d_x}\rightarrow\mathbb{R}^{d_x\times d_x}$ is
such that 
$$
\tilde \Sigma(s_2,x'):= \tilde \sigma(s_2,x') \tilde \sigma(s_2,x')^{\top} \equiv \Sigma(x').
$$ 
\eqref{eq:aux_SDE_tilde} is specified so that its transition density $\tilde{f}$ is available; see \cite[Section 2.2]{schauer} for the technical conditions on $\tilde \mu, \tilde \Sigma, \tilde f$. 
The main purpose of $\{\tilde X_t\}_{t\in[s_1,s_2]}$ is to sample $x'$ and use its transition density to construct another process $\{X_t^\circ\}_{t\in[s_1,s_2]}$ conditioned to hit $x'$ at $t=s_2$; which in turn will be an importance proposal for $\{X_t\}_{t\in[s_1,s_2]}$.  Set:
\begin{align}
\label{eq:aux_SDE}
d X^\circ_{t} = \mu_{s_2}^{\circ}(t,X^{\circ}_{t}; x')dt + \sigma(X^{\circ}_{t})dW_{t}, \quad t\in[s_1,s_2],\quad~X^{\circ}_{s_1} =x, 
\end{align}
where:
$$
\mu_{s_2}^{\circ}(t,x;x')=\mu(x)+\Sigma(x)\nabla_x\log\tilde{f}_{t,s_2}(x'|x),
$$ 
and denote by
 $\mathbb{P}^\circ_{x,x'}$ the probability law of the solution of (\ref{eq:aux_SDE}). 
The SDE in \eqref{eq:aux_SDE} yields: 
\begin{equation}
\mathbf{W}\rightarrow C_{s_1,s_2}(x,\mathbf{W}_{[s_1,s_2]},x'),
\label{eq:map}
\end{equation} 
mapping the driving Wiener noise $\mathbf{W}$ to the solution of \eqref{eq:aux_SDE}, reparametering the problem from $\mathbf{X}$ to $(\mathbf{W},x')$.

The authors in \cite{schauer} prove that $\mathbb{P}_{x,x'}$ and $\mathbb{P}^\circ_{x,x'}$ are absolutely continuous w.r.t.~each other, with Radon-Nikodym derivative: 
\begin{equation}
\frac{d\mathbb{P}_{x,x'} }{d \mathbb{P}^\circ_{x,x'} }(\mathbf{X}_{[s_1,s_2]})=
\exp\Big\{  \int_{s_1}^{s_2} L_{s_2}(t,X_t)dt\Big\}  \frac{ \tilde{f}_{s_1,s_2}(x'|x)}{f_{s_1,s_2}(x'|x)},
\label{eq:density}
\end{equation}
where: 
\begin{align*}
L_{s_2}&(t,x):=\left(\mu(x)- \tilde{\mu}(t,x)\right)^{\top}\, \nabla_x\log\tilde{f}_{t,s_2}(x'|x)\\
&-\frac{1}{2}\textrm{Tr}\,\Big\{\,\big[ \Sigma(x)-\tilde{\Sigma}(t,x)\big] \big[ -\nabla_x^2\log\tilde{f}_{t,s_2}(x'|x)-\nabla_x\log\tilde{f}_{t,s_2}(x'|x)\nabla_x\log\tilde{f}_{t,s_2}(x'|x)^{\top} \big]\,\Big\} ,         
\end{align*}
with $\textrm{Tr}(\cdot)$ denoting the trace of a squared matrix. 

We also need to include the sampling of $x'$, so set 
$$
R_{s_1,s_2}(\mathbf{X}_{[s_1,s_2]}) := \frac{d\mathbb{P}_{x,x'} }{d \mathbb{P}^\circ_{x,x'} }(\mathbf{X}_{[s_1,s_2]})
f_{s_1,s_2}(x'|x),
$$
where one should note that remarkably $R_{s_1,s_2}$ does not depend on the unknown transition density $f_{s_1,s_2}$.



\subsection{Bridges and Manifolds}

We will consider a linear observation process:
$$
Y_n  = A_n X_n + \Delta^{1/2}\epsilon_n 
$$
where $Y_n\in\mathbb{R}^{d_y}$,  $1\leq d_y < d_x$,  $A_n\in\mathbb{R}^{d_y\times d_x}$ is of full row rank, 
and independently for each $n\in\mathbb{N}$  $\epsilon_n\in\mathbb{R}^{d_y}$ are a sequence of independent noises with positive Lebesgue density $p_n$
and $\Delta\geq 0$. 
This is the same linear set-up as the previous section. Note that we are considering hidden state that evolves in continuous time $t$ that is observed at uniformly spaced discrete times and in particular when $t=n=1,2,3,\ldots$, i.e. at unit time intervals. This allows to simplify the notation and make it consistent with previous articles \cite{beskos}.  However,  this need not be the case and observations could in principle have a different amount of time in-between them,  or indeed be irregularly observed.

\subsubsection{The Low Noise Case}\label{sec:ln_diff}

Let $\mathbb{W}$ is the law of Brownian motion.
The probability measure associated to the smoother in continuous-time at time $n$ will be proportional to
\begin{equation}\label{eq:smooth_diff}
\prod_{k=1}^n\Big\{ 
R_{k-1,k}(C_{k-1,k}(u_{k-1}^{\Delta}(\widetilde{z}_{k-1},x),\mathbf{w}_{[k-1,k]},u_k^{\Delta}(\widetilde{z}_k,x)))
\mathbb{W}(d\mathbf{w}_{[k-1,k]})
p_k(u_k^{\Delta}(\widetilde{z}_k,\epsilon))\lambda_{d_x}(d\widetilde{z}_k)
\Big\}.
\end{equation}
In practice we cannot work with this and consider a time discretization. We first give the standard Euler-Maruyama time discretization of the solution to \eqref{eq:aux_SDE} (associated to a time interval $[k-1,k]$, $k\in\{1,\dots\}$) on a regular grid of spacing $\Xi_{l}=2^{-l}$, with starting point $x_{k-1}^{\circ}$ and ending point $x_{k}^{\circ}$.  The choice $\Xi_{l}=2^{-l}$, is again for consistency of presentation with \cite{beskos},  but other schemes can be used in practice.
We have,  for $j\in\{0,1\dots,\Xi_{l}^{-1}-2\}$:
\begin{align}\label{eq:disc_circ}
X_{k-1+(j+1)\Xi_{l}}^{\circ} & = X_{k-1+j\Xi_{l}}^{\circ} + 
\mu_{k}^{\circ}(k-1+j\Xi_{l},X^{\circ}_{k-1+j\Xi_{l}}; x_{k}^\circ)\Xi_{l} + 
\nonumber \\ &\qquad\qquad\qquad
\sigma(X^{\circ}_{k-1+j\Xi_{l}})\left[W_{k-1+(j+1)\Xi_{l}}-W_{k-1+j\Xi_{l}}\right].
\end{align}
Given $(x_{k-1}^{\circ},x_{k}^{\circ})$ and $\mathbf{W}_{[k-1,k]}^l=(W_{k-1+\Xi_{l}}-W_{k-1},\dots,W_{k-\Xi_l}-W_{k-2\Xi_{l}})$,  the recursion \eqref{eq:disc_circ} induces a discretized path 
$X_{k-1+\Xi_{l}}^{\circ},\dots,X_{k-\Xi_{l}}^{\circ}$ and we write such a path, including the starting and ending points with the notation
$$
C_{k-1,k}^l(x_{k-1}^{\circ},\mathbf{W}_{[k-1,k]}^l,x_{k}^{\circ}).
$$
We also need a discretization of the Radon-Nikodym derivative.  Consider 
$$
\mathbf{X}_{[k-1,k]}^l=(X_{k-1},X_{k-1+\Xi_{l}},\dots,X_k)
$$ 
then we set
$$
R_{k-1,k}^l(\mathbf{X}_{[k-1,k]}^l) := 
\exp\Big\{ \sum_{j=0}^{\Xi_{l}^{-1}-1} 
L_{k}(t,X_{k-1+j\Xi_{l}})\Xi_{l}\Big\}\tilde{f}_{k-1,k}(X_{k}|X_{k-1}).
$$

As a result of the time-discretization we can target the smoother that is proportional to
$$
\prod_{k=1}^n\Big\{ 
R_{k-1,k}^l(C_{k-1,k}^l(u_{k-1}^{\Delta}(\widetilde{z}_{k-1},x),\mathbf{w}_{[k-1,k]}^l,u_k^{\Delta}(\widetilde{z}_k,x)))
\mathbb{W}(d\mathbf{w}_{[k-1,k]}^l)
p_k(u_k^{\Delta}(\widetilde{z}_k,\epsilon))\lambda_{d_x}(d\widetilde{z}_k)
\Big\}
$$
For the importance sampling part, we proceed similar to Section \ref{sec:ln_add} and for the weights set:
\begin{align*}
w_n^{\Delta,l}(\widetilde{z}_{n-1},\mathbf{w}_{[n-1,n]}^l,\widetilde{z}_n)&  = 
R_{n-1,n}^l(C_{n-1,n}^l(u_{n-1}^{\Delta}(\widetilde{z}_{n-1},x),\mathbf{w}_{[n-1,n]}^l,u_n^{\Delta}(\widetilde{z}_n,x)))
\frac{
p_n(u_n^{\Delta}(\widetilde{z}_n,\epsilon))}
{\widetilde{q}_n^{\Delta}(\widetilde{z}_n|\widetilde{z}_{n-1})}.
\end{align*}

The procedural steps of the low-noise diffusion particle filter follow the generic diffusion SIR structure given in Algorithm~\ref{alg:diff_core} in Section~\ref{sec:diff_core}, with the proposal, weight, and state coordinate specified for the low-noise variant in Section~\ref{sec:generic_pf}.

\subsubsection{Degenerate Noise}\label{sec:dn_diff}

In the degenerate noise case,  one can follow the same path as the low noise case and write the time discretized smoother
up to a normalizing constant as
$$
\prod_{k=1}^n\Big\{ 
R_{k-1,k}^l(C_{k-1,k}^l(u_{k-1}^{\star}(z_{k-1}),\mathbf{w}_{[k-1,k]}^l,u_k^{\star}(z_k)))
\mathbb{W}(d\mathbf{w}_{[k-1,k]}^l)
\lambda_{d_x-d_y}(dz_k)
\Big\}.
$$
Similar to Section \ref{sec:dn_add}, for the importance weights we set:
\begin{align*}
w_n^{\star,l}(z_{n-1},\mathbf{w}_{[n-1,n]}^l,z_n)&  = 
\frac{R_{n-1,n}^l(C_{n-1,n}^l(u_{n-1}^{\star}(z_{n-1}),\mathbf{w}_{[n-1,n]}^l,u_n^{\star}(z_n)))
}{q_n^{\star}(z_n|z_{n-1})}.
\end{align*}

The procedural steps of the degenerate-noise diffusion particle filter follow the generic diffusion SIR structure given in Algorithm~\ref{alg:diff_core} in Section~\ref{sec:diff_core}, with the proposal, weight, and state coordinate specified for the degenerate variant in Section~\ref{sec:generic_pf}.

The algorithm here is presumably a PF implementation based on the ideas in \cite{bierkens} for the uniformly elliptic case. 
Note that the robustness to noise and discretization error could be proved in a similar manner to
Proposition \ref{res:2} and using similar assumptions to (A\ref{ass:1}). In more details one can envisage an unimplementable continuous time particle filter associated to our time discretized PF and formally prove that as $\Delta\downarrow 0$ and then $l\rightarrow\infty$ expectations of functions associated to the fixed $\Delta$ and $l$ PF converge to those of the continuous-time PF.   The latter proof is similar to Proposition \ref{res:2} and one would have to additionally prove results on the convergence of Euler-Maruyama time discretizations.

\subsubsection{Generic Diffusion PF Framework}\label{sec:diff_core}

For diffusion processes, the particle filters from Section~\ref{sec:generic_pf} are augmented with the bridge construction described in Section~\ref{sec:bridge}. Algorithm~\ref{alg:diff_core} gives the corresponding SIR structure with bridge correction. The four variants are obtained by using the same choices of $(q_n,w_n,Z_n)$ as in Algorithms~\ref{alg:bootstrap_pf}-\ref{alg:deg_pf}, and by multiplying the incremental weight by $R_{n-1,n}^l$ evaluated on the discretized bridge path.

\begin{algorithm}[ht]
\caption{Generic Diffusion PF Core}
\label{alg:diff_core}
\begin{algorithmic}[1]
\Require $N$, proposal $q_n$, weight $w_n$, data $\{y_n\}$, $x_0$, discretization level $l$ with $\Xi_l=2^{-l}$, bridge correction $R_{n-1,n}^l$.
\State Set $X_0^i=x_0$, $a_0^i=i$, $i=1,\ldots,N$, and $n=1$.
\While{new observation $y_n$}
    \State \textbf{Propagate:} Sample
    $$
        Z_n^i \sim q_n(\cdot \mid Z_{n-1}^{a_{n-1}^i}),
        \qquad i=1,\ldots,N,
    $$
    with $q_1(\cdot\mid x_0)$ used at the first step.
    \State Obtain $X_n^i$ from $Z_n^i$ as specified in Algorithms~\ref{alg:bootstrap_pf}-\ref{alg:deg_pf}.
    \State \textbf{Bridge:} Sample Brownian increments and generate
    $$
        \mathbf X_{[n-1,n]}^{i,l}
        =
        C_{n-1,n}^l
        \bigl(
        X_{n-1}^{a_{n-1}^i},
        W_{[n-1,n]}^{i,l},
        X_n^i
        \bigr),
        \qquad i=1,\ldots,N.
    $$
    \State \textbf{Weight:} Compute
    $$
        w_n^{l,i}
        =
        w_n\bigl(Z_{n-1}^{a_{n-1}^i},Z_n^i\bigr)
        R_{n-1,n}^l\bigl(\mathbf X_{[n-1,n]}^{i,l}\bigr),
        \qquad i=1,\ldots,N.
    $$
    \State \textbf{Normalize:}
    $$
        \rho_n^{l,i}
        =
        \frac{w_n^{l,i}}{\sum_{s=1}^N w_n^{l,s}},
        \qquad i=1,\ldots,N.
    $$
    \State \textbf{Estimate:}
    $$
        \pi_n^{N,l}(\varphi)
        =
        \sum_{i=1}^N \rho_n^{l,i}\,\varphi(X_n^i).
    $$
    \State \textbf{Resample:} Sample
    $$
        a_n^i \sim \mathrm{Cat}(\cdot;N,\{\rho_n^{l,s}\}_{s=1}^N),
        \qquad i=1,\ldots,N.
    $$
    \State Set $n \leftarrow n+1$.
\EndWhile
\end{algorithmic}
\end{algorithm}

%


\section{Numerical Results}\label{sec:numerics}

The four PF variants from Section~\ref{sec:generic_pf} are instantiated in the numerical experiments as follows. For the linear Gaussian model (Section~\ref{ex:lin_gauss}) and the nonlinear Lorenz-96 model with additive $t$-distributed noise (Section~\ref{ex:non_lin}), we use the closed-form optimal proposals and weights derived in Section~\ref{sec:spec_add} (with the $t$-distribution handled via a conditional scale-mixture representation). For the diffusion examples (Section~\ref{ex:diff}), the bridge construction of Section~\ref{sec:diff} is applied to all PF variants via the generic diffusion core (Algorithm~\ref{alg:diff_core}), with the auxiliary process chosen as a linearized Ornstein-Uhlenbeck process. In all cases, the generic SIR core (Algorithm~\ref{alg:sir_core}) remains unchanged; only the inputs to the core are modified.

For all numerical examples we have
\begin{equation*}
    Y_n = A_nX_n + \Delta^{1/2}\epsilon_n
\end{equation*}
where,  independently for each $n\in\mathbb{N}$,  $\epsilon_n\sim\mathcal{N}_{d_y}(0,\mathtt{I}_{d_y})$ and varying $\Delta$.   In Section \ref{ex:lin_gauss} $d_y=1$,  in Section \ref{ex:non_lin} $d_y=2$ and in Section \ref{ex:diff} $d_y=1$.

\subsection{Linear Gaussian Model}\label{ex:lin_gauss}

\begin{figure}[h]
    \centering
    \includegraphics[width=\linewidth]{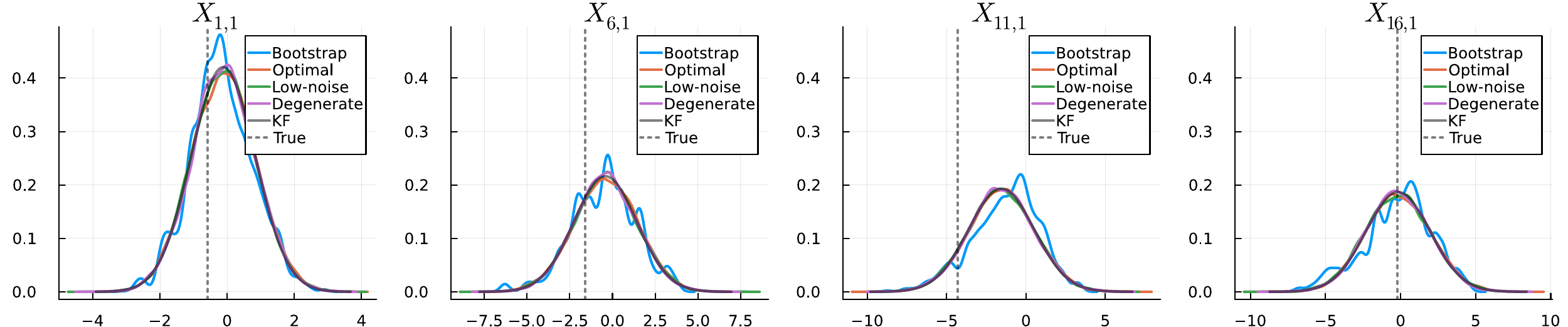}
    \caption{Marginal densities of $X_{n,1}$ for $n\in\{1,6,11,16\}$.  This is for the linear Gaussian model.}
    \label{fig:lgm_1}
\end{figure}

We consider a linear Gaussian model with $d_x=10$
\begin{equation*}
    X_n = B_nX_{n-1} + \nu_n
\end{equation*}
where,  independently for each $n\in\mathbb{N}$,  $\nu_n\sim\mathcal{N}_{d_x}(0,\Omega)$. We choose $X_0=(0,\dots,0)^\top$, $\Omega=\mathtt{I}_{d_x}$, $B_n=0.9\mathtt{I}_{d_x}$, $(A_n)_{ij}=1/d_x$ (computes the average of the state),  $n\in\{1,\dots,20\}$ and $\Delta=10^{-4}$. We run the Kalman filter, the particle filters with bootstrap and optimal proposals under the natural parameterization, and the proposed filters for the low noise and degenerate cases. We set the number of particles to $N=10^4$ and resample if the effective sample size (ESS) drops below $N/2$.  In Figure \ref{fig:lgm_1}, we plot the marginal densities of the $1$st component of $X_n$ at various $n$. We see that, except the bootstrap, all filters yield similar estimates as one might expect in this case. 

\begin{figure}[h]
    \centering
    \includegraphics[width=\linewidth]{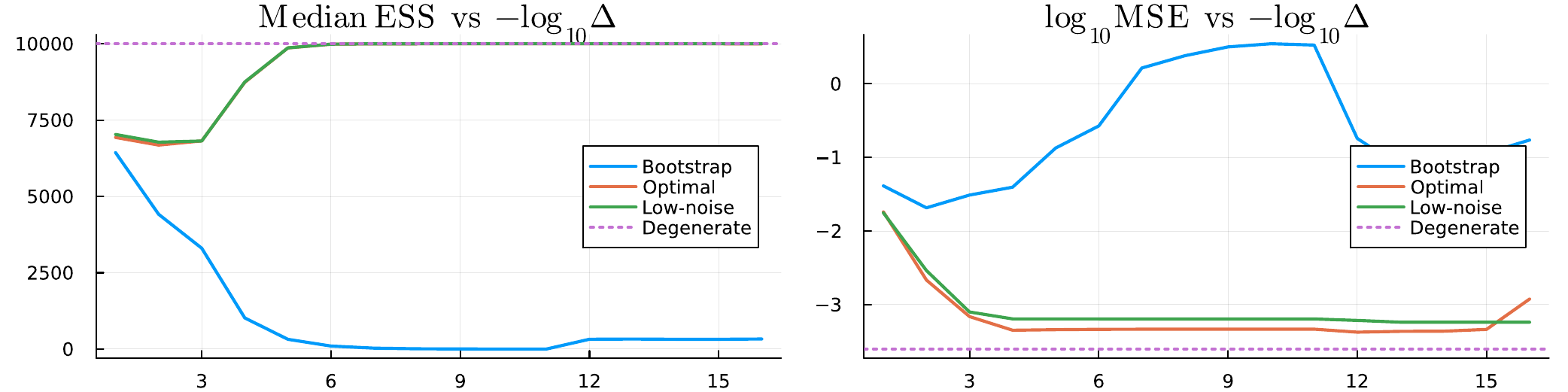}
    \caption{ESS (left) and MSE (right) versus $\Delta$.  This is for the linear Gaussian model.}
    \label{fig:lgm_3}
\end{figure}

We now repeat the experiment for various $\Delta$ and allow $\Delta\to0$.  We monitor the median ESS (Figure \ref{fig:lgm_3}) across the time-steps for a given $\Delta$.
In Figure \ref{fig:lgm_3}
we observe that the bootstrap PF degenerates for relatively large $\Delta$,  whilst the other methods remain stable. 
We can also see it in the mean squared error (MSE) of the mean estimates (error averaged across time and dimensions, see Equation \ref{eq:mse}), where the bootstrap PF collapses but the low-noise optimal proposal converges to the degenerate case.
\begin{equation}\label{eq:mse}
    \text{MSE} = \frac{1}{n_{\text{step}}}\sum_{n=1}^{n_{\text{step}}}\left\{ \frac{1}{d_x}\sum_{j=1}^{d_x} \left[ \left( \frac{1}{N}\sum_{i=1}^{N}X_{n,j}^{i} \right)-\hat{X}_{n,j}^{\mathrm{KF}} \right]^2 \right\}, \qquad \hat{X}_{n,j}^{\mathrm{KF}} = \mathbb{E}\left[X_{n,j}\mid Y_{1:n}\right].
\end{equation}

To assess the effect of the observation dimension, we fixed $d_x=10$ and varied $d_y$ from 1 to 9. We set $A_n$ to be the first $d_y$ rows of an orthogonal matrix $Q$ from the QR decomposition of a random $10\times 10$ matrix with i.i.d. standard normal entries generated using fixed seed. This gives $A_nA_n^\top = I_{d_y}$ and keeps the observation operator well conditioned for any $d_y$. The rest of the setup is as before: $X_0=(0,\dots,0)^\top$, $\Omega=I_{d_x}$, $B_n=0.9I_{d_x}$, $\Delta=10^{-4}$, $n=1,\dots,20$, $N=10^4$ particles, and resampling when ESS drops below $N/2$.

Table~\ref{tab:dy_experiment} reports the median ESS over iterations and MSE of the estimated mean, computed relative to the Kalman filter. The bootstrap PF collapses rapidly as we increase $d_y$; we report $0.00$ when the numerical underflow/instability prevents reliable weight normalization. In constrast, the low-noise and degenerate manifold PFs maintain ESS close to $N$ and low MSE for all values of $d_y$. The optimal proposal PF in natural parameterization also behaves well in this example. This experiment confirms that the stability of manifold methods is not due to low $d_y$; the methods remain effective as the codimension of the observation constraint increases.

\begin{table}[htbp]
\centering
\caption{Median ESS and MSE vs. observation dimension $d_y$ for the linear Gaussian model with $d_x=10$.}
\label{tab:dy_experiment}
\small
\begin{tabular}{lrrrrrrrr}
\hline
 & \multicolumn{2}{c}{Bootstrap} & \multicolumn{2}{c}{Optimal} & \multicolumn{2}{c}{Low-noise} & \multicolumn{2}{c}{Degenerate} \\
\cline{2-3} \cline{4-5} \cline{6-7} \cline{8-9}
$d_y$ & ESS & MSE & ESS & MSE & ESS & MSE & ESS & MSE \\
\hline
1 & 110.39 & 1.61e-01 & 9994.98 & 4.93e-04 & 9994.95 & 3.06e-04 & 10000.00 & 2.39e-04 \\
2 & 1.87 & 2.35e+00 & 9991.67 & 4.80e-04 & 9991.83 & 3.67e-04 & 10000.00 & 2.74e-04 \\
3 & 1.00 & 3.24e+00 & 9985.23 & 3.18e-04 & 9985.19 & 2.28e-04 & 10000.00 & 2.23e-04 \\
4 & 0.00 & 1.18e+00 & 9975.88 & 4.67e-04 & 9975.98 & 1.26e-04 & 10000.00 & 2.09e-04 \\
5 & 0.00 & 1.47e+00 & 9973.63 & 1.69e-04 & 9974.00 & 2.38e-04 & 10000.00 & 3.17e-04 \\
6 & 0.00 & 2.20e+00 & 9968.12 & 1.79e-04 & 9967.97 & 1.50e-04 & 10000.00 & 1.18e-04 \\
7 & 0.00 & 2.50e+00 & 9963.58 & 1.21e-04 & 9963.68 & 4.45e-05 & 10000.00 & 6.86e-05 \\
8 & 0.00 & 2.59e+00 & 9958.48 & 1.57e-04 & 9957.80 & 1.17e-04 & 10000.00 & 5.27e-05 \\
9 & 0.00 & 2.90e+00 & 9951.96 & 4.91e-05 & 9951.49 & 2.42e-05 & 10000.00 & 8.54e-05 \\
\hline
\end{tabular}
\end{table}

\subsection{Nonlinear Model with Additive Noise}\label{ex:non_lin}

\begin{figure}[h]
    \centering
    \includegraphics[width=\linewidth]{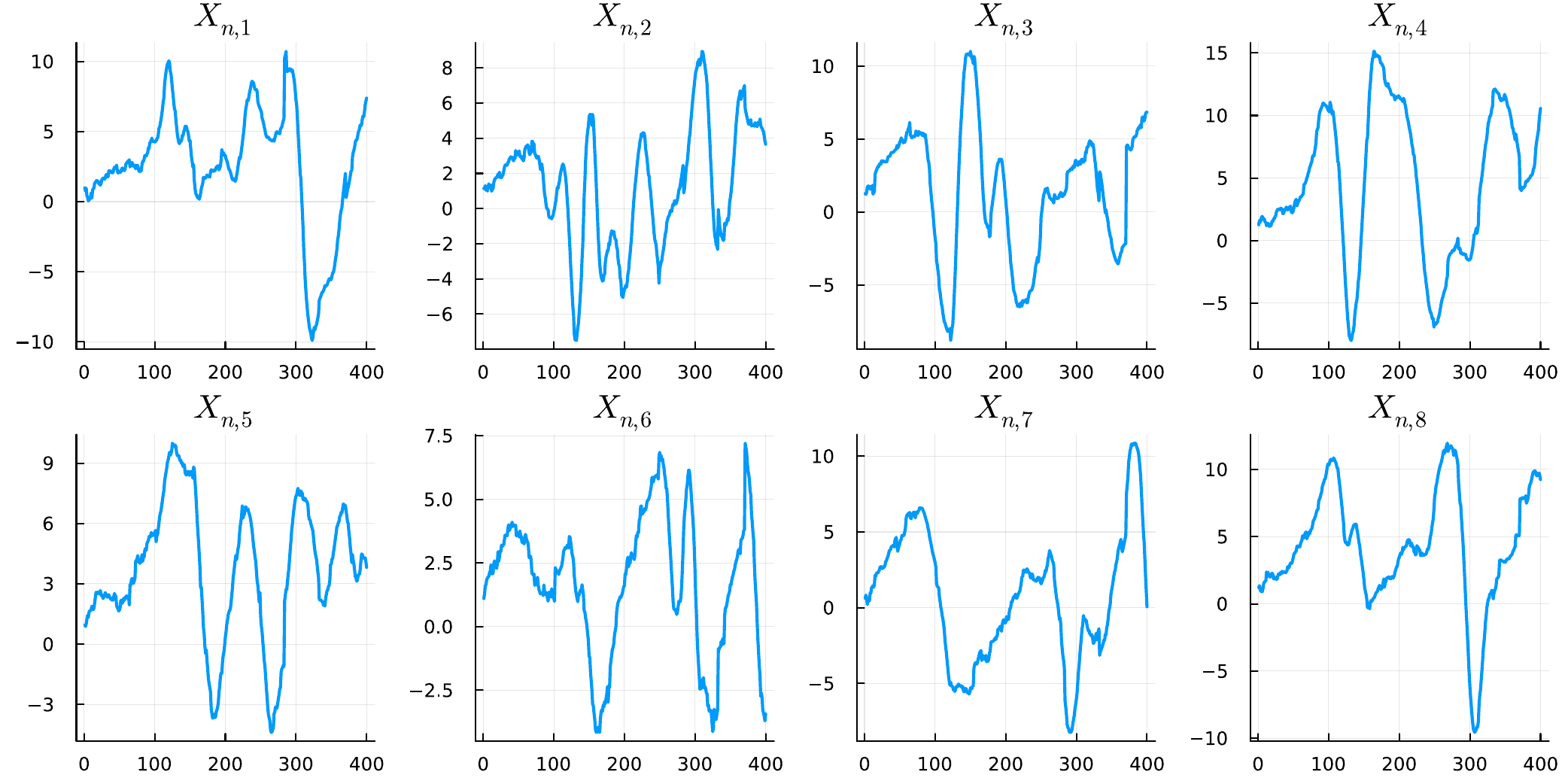}
    \caption{Components of $X_n$ for $n\in\{1,\dots,400\}$. This is for the nonlinear model.}
    \label{fig:lorenz_0}
\end{figure}

Now we consider a nonlinear Lorenz-96 model with additive $t$-distributed noise, $d_x=8$
\begin{equation*}
    X_n = F(X_{n-1}) + \nu_n
\end{equation*}
where $F_j(x) = x_j + \Delta t ((x_{j+1}-x_{j-2})x_{j-1}-x_j+F_0)$, $j\in\{1,\dots,d_x\}$ and,  independently for each $n\in\mathbb{N}$, $\nu_n\sim t_{d_x}(0,\Omega,\vartheta)$ (multivariate $t$-distribution with scale matrix $\Omega$ and $\vartheta>0$ degrees of freedom). We fix the parameter of the model to $F_0=8$ and the degrees of freedom of the innovation noise to $\vartheta=2$, and simulate it at stepsize $\Delta t=10^{-2}$ for $n\in\{1,\dots,400\}$ to obtain the data, starting from $X_0=(1,\dots,1)^\top$.  We fix $\Delta=10^{-4}$ and $\Omega=\Delta t\mathtt{I}_{d_x}$, the matrix $A_n$ is constructed such that we observe the $1$st and $5$th component of $X_n$. We can observe the chaotic behavior of the model in Figure \ref{fig:lorenz_0},  which are the simulated dynamics of the model.

Note that the optimal proposal is not available directly due to $\nu_n$ being $t$-distributed. However, it admits a hierarchical representation as a scale-mixture of Gaussians, via $\nu_n|\omega_n\sim\mathcal{N}(0,\Omega/\omega_n)$ with a Gamma-distributed auxiliary variable $\omega_n\sim\Gamma(\vartheta/2,\vartheta/2)$. Given $\omega_n$ the noise is conditionally Gaussian, so can use the optimal proposals from the Gaussian case to obtain a \emph{conditionally} optimal proposal for both the natural and noise parameterizations, conditioned on $\omega_n$, which we sample from the prior $\Gamma(\vartheta/2,\vartheta/2)$. 

\begin{figure}[h]
    \centering
    \includegraphics[width=\linewidth]{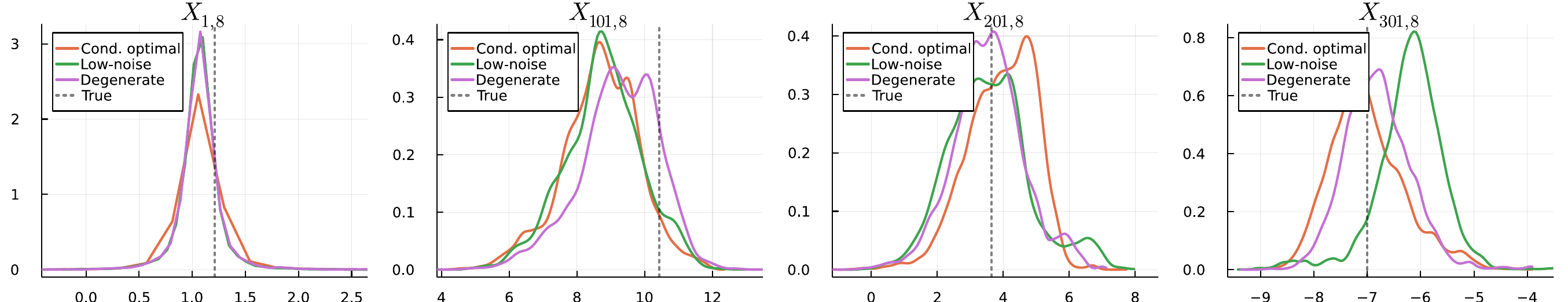}
    \caption{Marginal densities of $X_{n,8}$ for $n\in\{1,101,201,301\}$. This is for the nonlinear model.}
    \label{fig:lorenz_1}
\end{figure}

\begin{figure}[h]
    \centering
    \includegraphics[width=0.5\linewidth]{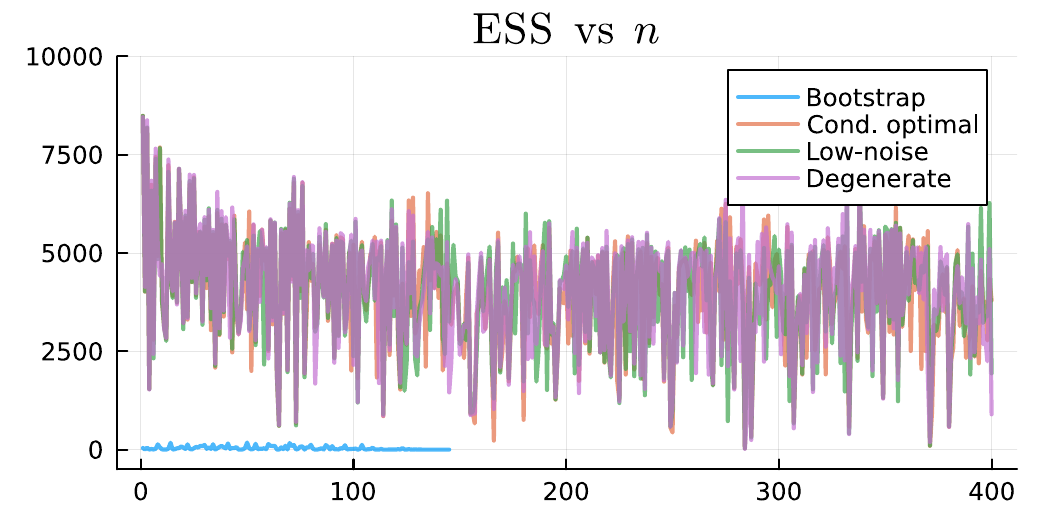}
    \caption{ESS versus $n$.  This is for the nonlinear model.}
    \label{fig:lorenz_2}
\end{figure}

\begin{figure}[h]
    \centering
    \includegraphics[width=\linewidth]{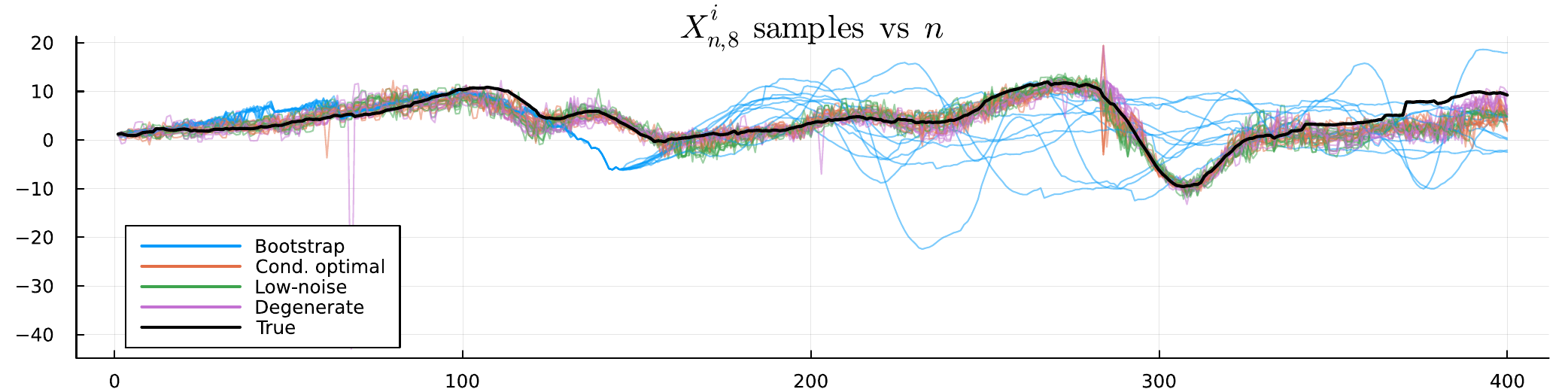}
    \includegraphics[width=\linewidth]{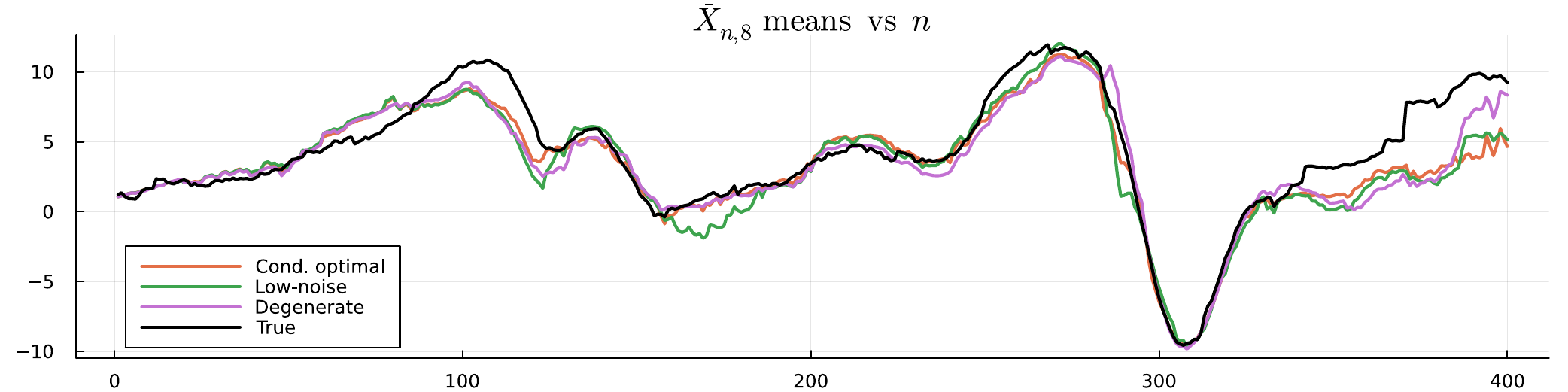}
    \caption{Sample paths (top) and mean of the filter (bottom) for $X_{n,8}$.  This is for the nonlinear model.}
    \label{fig:lorenz_3}
\end{figure}

\begin{figure}[h]
    \centering
    \includegraphics[width=\linewidth]{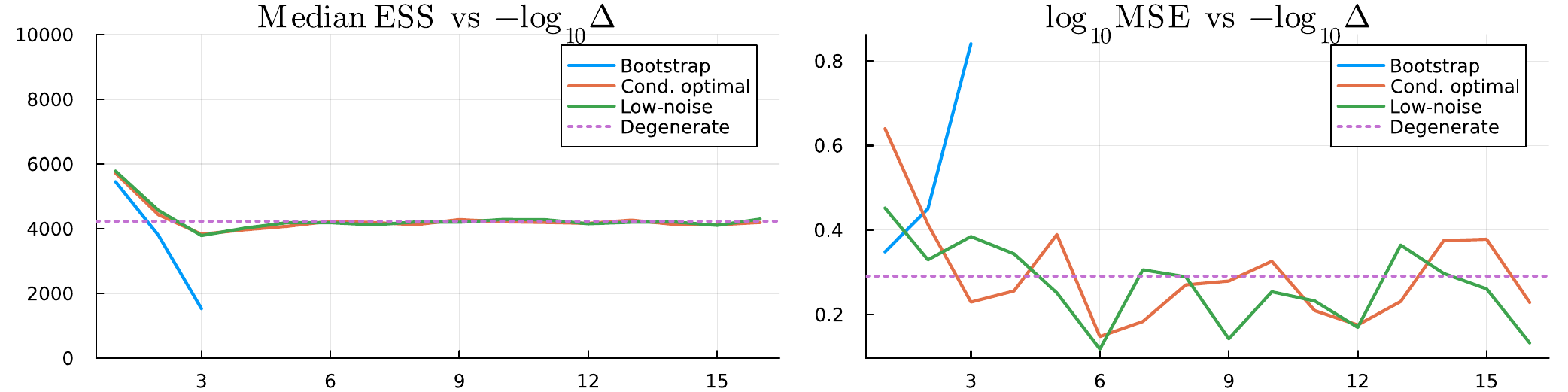}
    \caption{ESS (left) and MSE (right) versus $\Delta$. This is for the nonlinear model.}
    \label{fig:lorenz_5}
\end{figure}

We run all the mentioned filters (except for Kalman filter which does not apply here) again for $N=10^4$ particles with the same resampling rule. We plot several marginals in Figure \ref{fig:lorenz_1}, where we recover similar results as in the case of the linear Gaussian model, except for the case of the bootstrap PF whose performance is very poor. During later iterations, bootstrap weights collapse to zero, causing numerical issues and preventing us from producing any sensible plots. Further, we plot the ESS versus the time index in Figure \ref{fig:lorenz_2}, where we see that the bootstrap PF  collapses from the very beginning, while other methods give adequate results in terms of this measure.
To further examine the accuracy of the PFs we plot individual sample paths of the particles (for a given run) and their mean for $8$th component in Figure \ref{fig:lorenz_3}.  In both the top and bottom of  Figure \ref{fig:lorenz_3} the true value is the actual signal used to generate the data.   Apart from the bootstrap PF all of the methods are seemingly able to recover the true hidden state in the $8$th component.
In Figure \ref{fig:lorenz_5} we consider re-running all of the methods for different $\Delta$ and sending $\Delta\to 0$. 
Figure \ref{fig:lorenz_5} displays the median ESS (across time) and the MSE (between estimated mean and the true signal $\hat{X}_{n,j}^{\mathrm{true}}$) of the four PFs.
In this case the optimal proposal under the natural parameterization does not seem to collapse (in terms of ESS)  but this could be as the weight does not degenerate to zero. The MSE of the bootstrap is too large as it collapses immediately. The other two seem to roughly approach the MSE of the degenerate PF, but there is still variation due to iterates not being exactly the same.

\subsection{Diffusion Model}\label{ex:diff}

\begin{figure}[h]
    \centering
    \includegraphics[width=\linewidth]{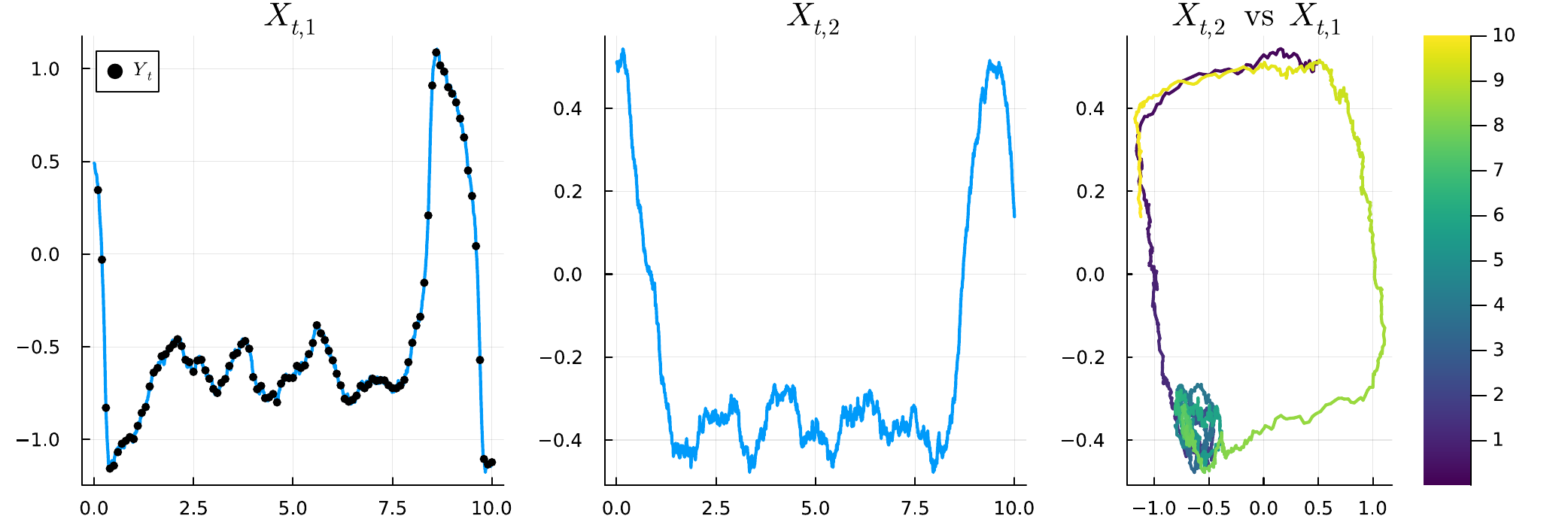}
    \caption{Components of $X_t$ for $t\in[0,10]$. This is for the diffusion model.}
    \label{fig:fhn_0}
\end{figure}

We now consider the FitzHugh-Nagumo diffusion model with $d_x=2$:
\begin{equation*}
    dX_t = \mu(X_t)dt + \sigma(X_t)dW_t   
\end{equation*}
with $\mu(x)=((x_1-x_1^3-x_2)/\alpha, \gamma x_1-x_2+\beta)^\top$ and $\sigma(x)=\sigma_0 \mathtt{I}_{d_x}$. This is an elliptic diffusion with constant diffusion term.  We fix the model parameters to $\alpha=0.1$, $\gamma=1$, $\beta=0.2$ and $\sigma_0=0.1$. We choose the starting values to be $X_0=(0.5, 0.5)^\top$. We simulate the dynamics in the interval $t\in[0,10]$ using the Euler-Maruyama scheme with stepsize $\Xi=5\times10^{-3}$ and observe the $1$st component at regular times with spacing $10^{-1}$ and finally $\Delta=10^{-8}$ is the noise level. This scenario leads to 20 discretization points between two consecutive observations and 2000 in total. The simulated dynamics is given in Figure \ref{fig:fhn_0}.

We consider an auxiliary process with linearized drift $\tilde{\mu}(x)=\mu(x_{\text{ref}})+\frac{\partial}{\partial x}\mu(x_{\text{ref}})(x-x_{\text{ref}})$.  We can calculate the reference point $x_{\text{ref}}$ by evolving the true dynamics deterministically starting from $x$ using the stepsize $\Xi$ for 20 iterations until we reach the next observation time. 
The auxiliary process is an Ornstein-Uhlenbeck process with a transition density of $x'$ given $x$ over time step $h$ as $\mathcal{N}_2(m, Q)$ where
\begin{equation}\label{eq:diff_aux_transition}
\begin{aligned}
    & m=\exp(Jh)x+J^{-1}(\exp(Jh)-I)(\mu-Jx_{\text{ref}}) \\
    & Q=\exp(Jh)\Psi\exp(J^\top h)-\Psi
\end{aligned}
\end{equation}
where $J=\frac{\partial}{\partial x}\mu(x_{\text{ref}})$, $\mu=\mu(x_{\text{ref}})$ and $\text{vec}(\Psi)=(I\otimes J+J\otimes I)^{-1}\text{vec}(\Sigma)$, $\otimes$ is a Kronecker product and $\exp(Jh)$ is a matrix exponentiation. Here, we assume that the jacobian $J$ is both diagonalizable and invertible.

We consider 4 PFs, the low noise and degenerate cases have been detailed. We also consider a `bootstrap' and `optimal' PF which are a little different from the conventional ones we have used so far.  Both the bootstrap and
optimal methods consider a smoother of the type \eqref{eq:smooth_diff} except under the natural parameterization.
That is, after time discretization,  the target smoother is up to a constant at time $n$
$$
\prod_{k=1}^n\Big\{ h_k(y_k|x_k)
R_{k-1,k}^l(C_{k-1,k}^l(x_{k-1},\mathbf{w}_{[k-1,k]}^l,
x_k))
\mathbb{W}(d\mathbf{w}_{[k-1,k]}^l)
\lambda_{d_x}(x_k)
\Big\}.
$$
The bootstrap PF will use a proposal that is exactly the auxiliary process (on $x_k$ the Brownian motion is sampled from its law)  whereas the optimal (which is not optimal in any sense) is a proposal proportional to the term $h_k(y_k|x_k)$ multiplied by the density of the auxiliary process given by Equation \ref{eq:diff_aux_transition} and the Brownian motion sampled from its law.

\begin{figure}[h]
    \centering
    \includegraphics[width=0.5\linewidth]{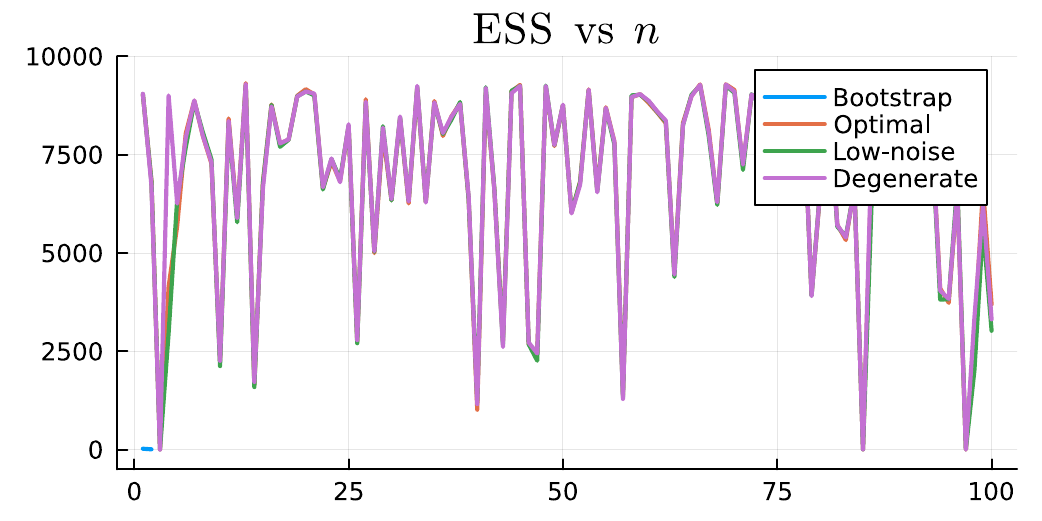}
    \caption{ESS versus $n$.  This is for the diffusion model.}
    \label{fig:fhn_2}
\end{figure}

\begin{figure}[h]
    \centering
    \includegraphics[width=\linewidth]{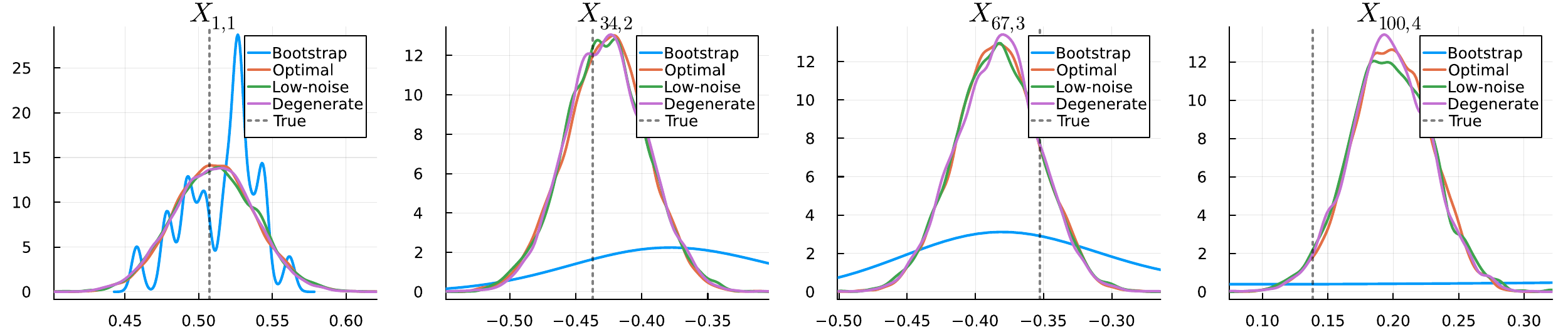}
    \caption{Marginals of $X_{n,2}$ for $n\in\{1, 34, 67, 100\}$.  This is for the diffusion model.}
    \label{fig:fhn_1}
\end{figure}

\begin{figure}[h]
    \centering
    \includegraphics[width=\linewidth]{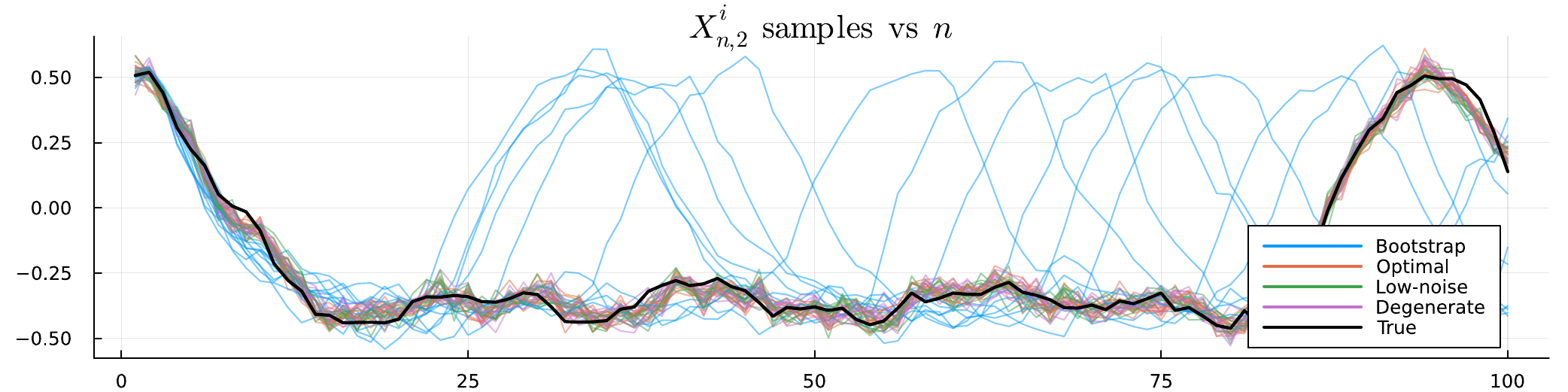}
    \includegraphics[width=\linewidth]{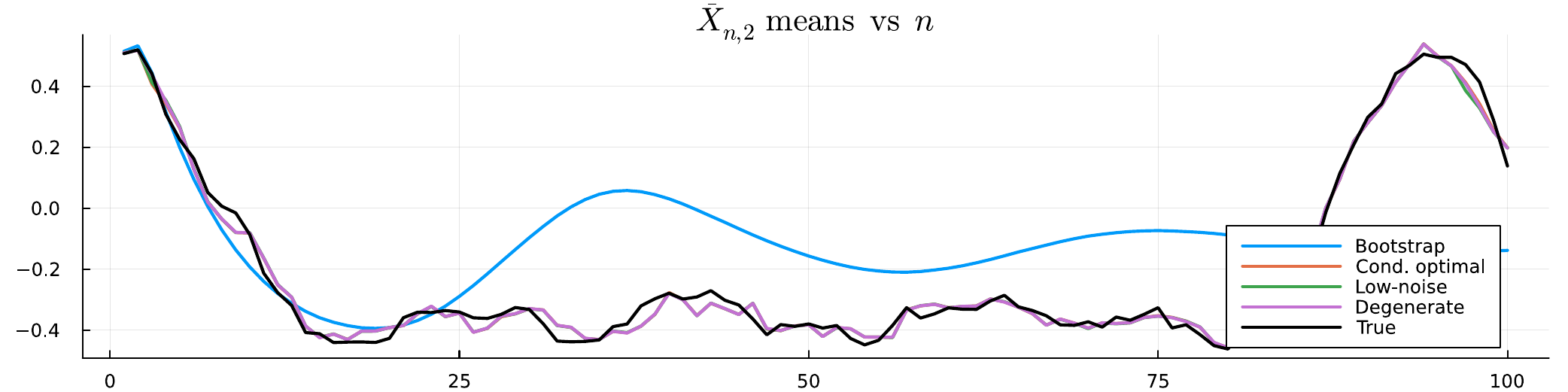}
    \caption{Sample paths (top) and mean estimates (bottom) versus time. This is for the diffusion model.}
    \label{fig:fhn_3}
\end{figure}

\begin{figure}[h]
    \centering
    \includegraphics[width=\linewidth]{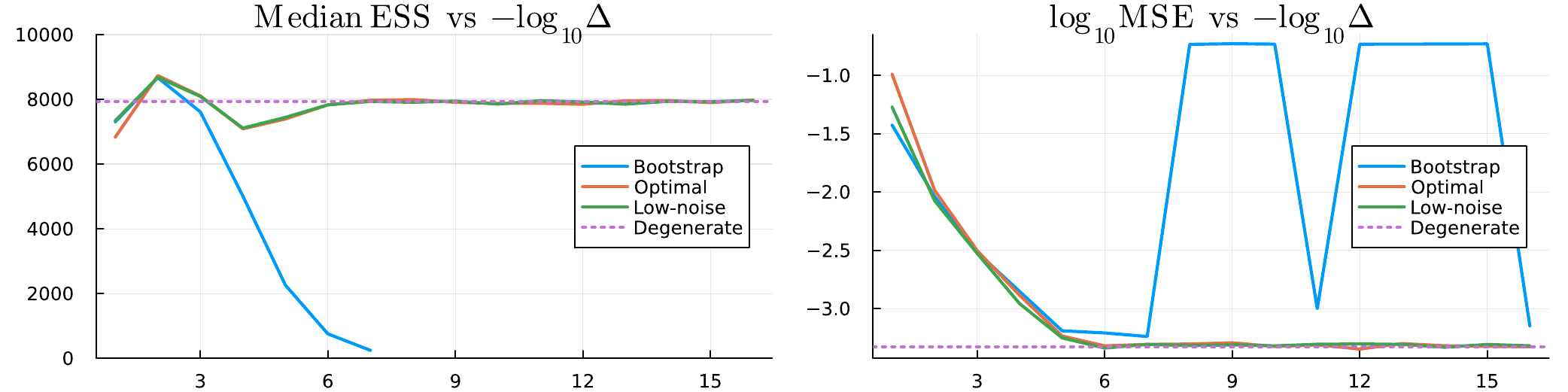}
    \caption{ESS (left) and MSE (right) versus $\Delta$. This is for the diffusion model.}
    \label{fig:fhn_5}
\end{figure}

In Figure \ref{fig:fhn_2} we plot the ESS against time for a particular run. Here, we have set $N=10^4$ and resample at each step.  As in the previous example, the bootstrap PF collapses completely and the other filters perform quite well for this criterion.  This shows that one can filter in low or degenerate noise and with a very fine time discretization.  This deduction is supported by the results in Figures \ref{fig:fhn_1} (marginal densities of the state) and \ref{fig:fhn_3} (tracking the true signal) which shows that one can reconstruct the signal process in this case. In Figure \ref{fig:fhn_5} we repeat the experiment for $\Delta\to0$, where we see how the bootstrap PF collapses and the low-noise PF converges to the degenerate case.

\subsubsection{State-dependent Diffusion Model}\label{ex:diff_2}

A more interesting example is when the diffusion term in FitzHugh-Nagumo model depends on the state.  We now revisit the previous model with exactly the same setting, except with the diffusion coefficient
\begin{equation*}
    \sigma(x) = \begin{bmatrix}
        \sigma_1\sqrt{x_1^2+1} & 0 \\  0 & \sigma_2\sqrt{x_2^2+1}
    \end{bmatrix}
\end{equation*}
where we fix $\sigma_1=\sigma_2=0.1$.  We remark that,  contrary to the previous example (constant diffusion coefficient),  we do not know of any potential alternative PF method.  In the constant diffusion case,  one could envisage using the Girsanov method w.r.t.~a linear diffusion with the same constant drift,  but that is not possible in this case.  The auxiliary process is as before except with $\tilde{\sigma}=\sigma(x')$, $x'$ the end-point. 
To actually propose the new position of the particle instead of using the auxiliary process, 
we use an approximation of the auxiliary density by substituting $x'$ with deterministic $x_{\text{ref}}$ and using $\sigma(x_{\text{ref}})$.   This incurs an extra correction term in the importance weights as now one needs to divide by the associated density.  The `bootstrap' and `optimal' methods use this new density in lieu of the auxiliary process density used in the constant diffusion case.  We run each PF with $N=10^4$ particles.

\begin{figure}[h]
    \centering
    \includegraphics[width=\linewidth]{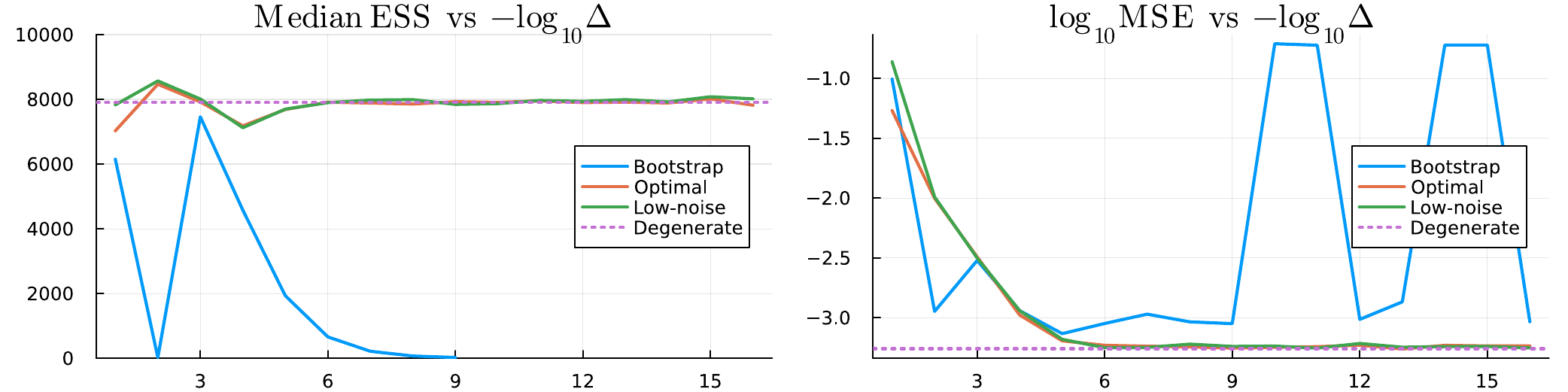}
    \caption{ESS (left) and MSE (right) versus $\Delta$. This is for the state-dependent diffusion model.}
    \label{fig:fhndiff_5}
\end{figure}

We consider the ESS versus $\Delta$ in Figure \ref{fig:fhndiff_5}.  We observe similar results to the case of a constant diffusion coefficient.   This shows that one can use the method for non-constant diffusions in low noise and high time discretization error.

\section{Conclusions}\label{sec:conc}

We have presented a novel PF framework that can be very effective for likelihoods with low and even degenerate observational noise. We have focused on the linear observation case and our numerical results showed our method is very accurate in a variety of settings. Our method takes advantage of being able to compute $V_n$ or $V_n^\Delta$ and having an explicit maps $u_k^\star$ or $u_k^\Delta$, so extending this PF methodology for the general non-linear manifolds whilst challenging would be interesting future work. Finally, the presented up reparameterizations should be also useful for Sequential Monte Carlo methods applied to Approximate Bayesian Computation and interpreting $\Delta$ as a tolerance threshold. 

In comparison to the work of \cite{zhu},  which is based on sequential MCMC (SMCMC) 
and focusses on the same problem in more generality, 
we have the following remarks,  which expand upon those in the introduction.
\textcolor{blue}{
SMCMC and PF are fundamentally different algorithms, and the PF-specific contributions of this work---the reparameterized recursion, optimal proposals, convergence theory, and diffusion extension---do not appear in \cite{zhu}. The manifold construction is a common tool, so it is no mystery why it looks similar, but the algorithmic developments and theoretical results are entirely distinct.
}

SMCMC will require,  in many cases, quite expensive computation as it relies on an empirical approximation of the current filter that is fed into (for example) the acceptance probability of an MCMC step.  If there are $N$ samples used to approximate the filter the cost is $\mathcal{O}(N^2)$ and although there is an auxiliary variable trick, we have found that tuning this variable takes a lot of work,  which may be less appropriate for filtering where one needs an online method.  In addition, \cite{zhu} does not address the problem of filtering diffusions and producing a method which is robust to discretization error and noise simultaneously.  That seems troublesome as the SMCMC method would need access to approximations of the transition density on $\mathcal{O}(1)$ time which would not degenerate with the discretization parameter.  Whilst it could be possible, that is not presented in \cite{zhu}.

\appendix

\section{Proofs and Auxiliary Material}\label{app:proof}

\subsection{On construction of $V_n^\Delta$}

\textcolor{blue}{
We choose
\begin{equation*}
V_n^\Delta =
\begin{pmatrix}
V_n & -\Delta^{1/2} A_n^\top (A_n A_n^\top)^{-1} \\
0 & I_{d_y}
\end{pmatrix}.
\end{equation*}
This matrix satisfies
$[A_n,\Delta^{1/2}I_{d_y}]\,V_n^\Delta = 0$,
so its columns span $\ker([A_n,\Delta^{1/2}I_{d_y}])$.  
Moreover, the corresponding noise coordinate is exactly the observation noise:
$u_n^\Delta(\widetilde z_n,\epsilon)=\overline{z}_n$.
The induced Gram matrix is
\begin{equation*}
(V_n^\Delta)^\top V_n^\Delta =
\begin{pmatrix}
I_{d_x-d_y} & 0\\
0 & I_{d_y}+\Delta (A_n A_n^\top)^{-1}
\end{pmatrix},
\end{equation*}
which is constant in $\widetilde z_n$ and therefore cancels in normalized importance weights and target densities.  
As $\Delta\downarrow0$, $V_n^\Delta\to \begin{pmatrix}V_n&0\\0&I_{d_y}\end{pmatrix}$, consistent with the degenerate case.  
In practice one may use any basis of the kernel, e.g.\ obtained by QR decomposition; the choice above is convenient for derivations.
}

\subsection{Proof of Proposition~\ref{res:2}}\label{app:res2}


\begin{proof}
\textcolor{blue}{
Suppose $\widetilde{z}_n^{\top} = (z_n^{\top},\overline{z}_n^{\top})^{\top}$,  with $(z_n,\overline{z}_n)\in \mathbb{R}^{d_x-d_y}\times \mathbb{R}^{d_y}$. Based on the above discussion  we have that for any fixed $\widetilde{z}_n\in\mathbb{R}^{d_x}$ 
\begin{equation}\label{eq:prf1}
u_n^{\Delta}(\widetilde{z}) \rightarrow
\begin{bmatrix}
x_n^{\star} \\
0
\end{bmatrix} + 
\begin{bmatrix}
V_n z_n  \\
\overline{z}_n
\end{bmatrix},\quad\text{as}\quad{\Delta\downarrow 0}.
\end{equation}
}
Then using \eqref{eq:prf1} and (A\ref{ass:1}) 2.~we have for any fixed $(\widetilde{z}_{k-1},\widetilde{z}_k)$ and any $k\in\{1,\dots,n\}$
\begin{equation}\label{eq:prf_lim}
\lim_{\Delta\downarrow 0}
p_k(u_k^{\Delta}(\widetilde{z}_k,\epsilon))
f_k(u_k^{\Delta}(\widetilde{z}_k,x)|u_{k-1}^{\Delta}(\widetilde{z}_{k-1},x))  = p_k(\overline{z_k})
f_k(u_k^{\star}(z_k)|u_{k-1}^{\star}(z_{k-1})).
\end{equation}
We remark that for any $\varphi\in\mathtt{B}_b(\mathbb{R}^{d_x-d_y})$
$$
\int_{\mathbb{R}^{nd_x}} \varphi(z_n)  \prod_{k=1}^n p_k(u_k^{\Delta}(\widetilde{z}_k,\epsilon))
f_k(u_k^{\Delta}(\widetilde{z}_k,x)|u_{k-1}^{\Delta}(\widetilde{z}_{k-1},x)) 
\bigotimes_{i=1}^n \lambda_{d_x}(d\widetilde{z}_i) = 
$$

$$
\int_{\mathbb{R}^{nd_x}} \varphi(z_n)  \left\{\prod_{k=1}^n w_k^{\Delta}(\widetilde{z}_{k-1},\widetilde{z}_{k})
\right\}
\prod_{k=1}^n \widetilde{q}_k^{\Delta}(\widetilde{z}_{k}|\widetilde{z}_{k-1})
\bigotimes_{i=1}^n \lambda_{d_x}(d\widetilde{z}_i).
$$
By (A\ref{ass:1}) 3.~\& 4.~we have that 
\begin{equation}\label{eq:prf_upper}
\left\{\prod_{k=1}^n w_k^{\Delta}(\widetilde{z}_{k-1},\widetilde{z}_{k})
\right\}
\prod_{k=1}^n \widetilde{q}_k^{\Delta}(\widetilde{z}_{k}|\widetilde{z}_{k-1}) \leq C
\prod_{k=1}^n p_k(\overline{z}_k)q_k^{\star}(z_k|z_{k-1})
\end{equation}
for any $\widetilde{z}_1,\dots,\widetilde{z}_n$ and $C$ a constant that does not depend on $\Delta$ nor
$\widetilde{z}_1,\dots,\widetilde{z}_n$.  As the R.H.S.~of \eqref{eq:prf_upper} is integrable 
and noting \eqref{eq:prf_lim}
one can apply the dominated convergence theorem. That is,  we have for any $\varphi\in\mathtt{B}_b(\mathbb{R}^{d_x-d_y})$
$$
\lim_{\Delta\downarrow 0} \int_{\mathbb{R}^{nd_x}} \varphi(z_n)  \prod_{k=1}^n p_k(u_k^{\Delta}(\widetilde{z}_k,\epsilon))
f_k(u_k^{\Delta}(\widetilde{z}_k,x)|u_{k-1}^{\Delta}(\widetilde{z}_{k-1},x)) 
\bigotimes_{i=1}^n \lambda_{d_x}(d\widetilde{z}_i)
 = 
$$
$$
\int_{\mathbb{R}^{nd_x}} \varphi(z_n)  \prod_{k=1}^n p_k(\overline{z}_k)
f_k(u_k^{\star}(z_k)|u_{k-1}^{\star}(z_{k-1})) 
\bigotimes_{i=1}^n \lambda_{d_x}(d\widetilde{z}_i) 
 =
$$
$$ 
 \int_{\mathbb{R}^{n(d_x-d_y)}}
\varphi(z_n)  \prod_{k=1}^n 
f_k(u_k^{\star}(z_k)|u_{k-1}^{\star}(z_{k-1})) 
\bigotimes_{i=1}^n \lambda_{d_x}(dz_i)
$$
and hence that
$$
\lim_{\Delta\downarrow 0}\pi_n^{\Delta}(\varphi) = \pi_n^{\star}(\varphi).
$$

Now,  using \eqref{eq:prf1} and assumption (A\ref{ass:1})  1.~\& 2.~we have for any fixed $(\widetilde{z}',\widetilde{z})\in\mathbb{R}^{d_x}$
\begin{equation}\label{eq:prf2}
\lim_{\Delta\downarrow 0} w_n^{\Delta}(\widetilde{z}',\widetilde{z}) = 
\frac{p_n(\overline{z})f_n(u_n^{\star}(z)|u_{n-1}^{\star}(z'))}
{p_n(\overline{z})q_n(z|z')} = w_n^{\star}(z',z).
\end{equation}
Now the joint density of the PF in the low noise case is
$$
\mathbb{P}^{\Delta}_n(\widetilde{z}_{1:n}^{1:N},\widetilde{a}_{1:n-1}^{1:N})
= \left\{\prod_{i=1}^N \widetilde{q}_1^{\Delta}(\widetilde{z}_1^i|x_0)\right\}\left\{
\prod_{k=2}^n \prod_{i=1}^N \frac{w_{k-1}^{\Delta}(\widetilde{z}_{k-2}^{\widetilde{a}_{k-2}^i},\widetilde{z}_{k-1}^{\widetilde{a}_{k-1}^i})}{
\sum_{s=1}^N
w_{k-1}^{\Delta}(\widetilde{z}_{k-2}^{\widetilde{a}_{k-2}^s},\widetilde{z}_{k-1}^{s})}
\widetilde{q}_k^{\Delta}(\widetilde{z}_k^i|\widetilde{z}_{k-1}^{\widetilde{a}_{k-1}^i})
\right\}.
$$
Therefore,  applying \eqref{eq:prf2} and (A1) 1.~for any fixed $\widetilde{z}_{1:n}^{1:N},\widetilde{a}_{1:n-1}^{1:N}$ we have that
$$
\lim_{\Delta\downarrow 0} \mathbb{P}^{\Delta}_n(\widetilde{z}_{1:n}^{1:N},\widetilde{a}_{1:n-1}^{1:N}) = 
$$
$$
\left\{\prod_{i=1}^N p_1(\overline{z}_1^i)q_1^{\star}(z_1^i|x_0)\right\}\left\{
\prod_{k=2}^n \prod_{i=1}^N \frac{w_{k-1}^{\star}(z_{k-2}^{\widetilde{a}_{k-2}^i},z_{k-1}^{\widetilde{a}_{k-1}^i})}{
\sum_{s=1}^N
w_{k-1}^{\star}(z_{k-2}^{\widetilde{a}_{k-2}^s},z_{k-1}^{s})}
p_k(\overline{z}_k^i) q_k^{\star}(z_k^i|z_{k-1}^{\widetilde{a}_{k-1}^i})
\right\}.
$$
We note also that by (A\ref{ass:1}) 3.
\begin{eqnarray*}
\mathbb{P}^{\Delta}_n(\widetilde{z}_{1:n}^{1:N},\widetilde{a}_{1:n-1}^{1:N}) & \leq &
\left\{\prod_{i=1}^N \widetilde{q}_1^{\Delta}(\widetilde{z}_1^i|x_0)\right\}\left\{
\prod_{k=2}^n \prod_{i=1}^N 
\widetilde{q}_k^{\Delta}(\widetilde{z}_k^i|\widetilde{z}_{k-1}^{\widetilde{a}_{k-1}^i})
\right\}\\
& \leq & 
C\left\{\prod_{i=1}^N p_1(\overline{z}_1^i)q_1^{\star}(z_1^i|x_0)\right\}\left\{
\prod_{k=2}^n \prod_{i=1}^N 
p_k(\overline{z}_k^i)q_k^{\star}(z_k^i|z_{k-1}^{\widetilde{a}_{k-1}^i})
\right\}
\end{eqnarray*}
where $C$ does not depend on $\Delta$ or any of the particle values.  The expression on the R.H.S.~of the above
inequality is integrable over $\{1,\dots,N\}^{n-1}\times\mathbb{R}^{Nnd_x}$, in that one produces a finite constant that is $\Delta$ independent.  We remark that 
$$
\left|\pi_n^{\Delta,N}(\varphi)-\pi_n^{\Delta}(\varphi)\right| \leq 2\sup_{z\in\mathbb{R}^{d_x-d_y}}|\varphi(z)|.
$$
As a result,  by the dominated convergence theorem we have that
$$
\lim_{\Delta \downarrow 0}\mathbb{E}^{\Delta}\left[\left|\pi_n^{\Delta,N}(\varphi)-\pi_n^{\Delta}(\varphi)\right|^r\right] = 
$$
$$
\sum_{\widetilde{\mathbf{a}}_{1:n-1}^{1:N}\in\{1,\dots,N\}^{n-1}} \int_{\mathbb{R}^{Nnd_x}}
\left|
\frac{\sum_{i=1}^N \varphi(z_n^i)
w_{n}^{\star}(z_{n-1}^{\widetilde{a}_{n-1}^i},z_{n}^{i})}
{\sum_{i=1}^N
w_{n}^{\star}(z_{n-1}^{\widetilde{a}_{n-1}^i},z_{n}^{i})}
-\pi_n^{\star}(\varphi)
\right|^{r} \times
$$
$$
\left\{\prod_{i=1}^N p_1(\overline{z}_1^i)q_1^{\star}(z_1^i|x_0)\right\}\left\{
\prod_{k=2}^n \prod_{i=1}^N \frac{w_{k-1}^{\star}(z_{k-2}^{\widetilde{a}_{k-2}^i},z_{k-1}^{\widetilde{a}_{k-1}^i})}{
\sum_{s=1}^N
w_{k-1}^{\star}(z_{k-2}^{\widetilde{a}_{k-2}^s},z_{k-1}^{s})}
p_k(\overline{z}_k^i) q_k^{\star}(z_k^i|z_{k-1}^{\widetilde{a}_{k-1}^i})
\right\}
\bigotimes_{k=1}^n\bigotimes_{i=1}^N \lambda_{d_x}(d\widetilde{z}^i_k) = 
$$
$$
\sum_{\widetilde{\mathbf{a}}_{1:n-1}^{1:N}\in\{1,\dots,N\}^{n-1}} \int_{\mathbb{R}^{Nn(d_x-d_y)}}
\left|
\frac{\sum_{i=1}^N \varphi(z_n^i)
w_{n}^{\star}(z_{n-1}^{\widetilde{a}_{n-1}^i},z_{n}^{i})}
{\sum_{i=1}^N
w_{n}^{\star}(z_{n-1}^{\widetilde{a}_{n-1}^i},z_{n}^{i})}
-\pi_n^{\star}(\varphi)
\right|^{r} \times
$$
$$
\left\{\prod_{i=1}^N q_1^{\star}(z_1^i|x_0)\right\}\left\{
\prod_{k=2}^n \prod_{i=1}^N \frac{w_{k-1}^{\star}(z_{k-2}^{\widetilde{a}_{k-2}^i},z_{k-1}^{\widetilde{a}_{k-1}^i})}{
\sum_{s=1}^N
w_{k-1}^{\star}(z_{k-2}^{\widetilde{a}_{k-2}^s},z_{k-1}^{s})}
 q_k^{\star}(z_k^i|z_{k-1}^{\widetilde{a}_{k-1}^i})
\right\}
\bigotimes_{k=1}^n\bigotimes_{i=1}^N \lambda_{d_x-d_y}(dz^i_k) 
$$
where $\widetilde{\mathbf{a}}_{1:n-1}^{1:N} = (\widetilde{a}_1^1,\dots,\widetilde{a}_1^N,\dots,\widetilde{a}_{n-1}^1,\dots,\widetilde{a}_{n-1}^N)$.  We recognize the last expression as the expectation w.r.t.~the law of the degenerate noise PF; i.e.~we have proved that
$$
\lim_{\Delta \downarrow 0}\mathbb{E}^{\Delta}\left[\left|\pi_n^{\Delta,N}(\varphi)-\pi_n^{\Delta}(\varphi)\right|^r\right]^{1/r} = 
\mathbb{E}^{\star}\left[\left|\pi_n^{\star,N}(\varphi)-\pi_n^{\star}(\varphi)\right|^r\right]^{1/r}.
$$
\end{proof}

\subsection{Proof of Proposition~\ref{prop:prop}}\label{app:inc}

\begin{proof}
The results under the natural parameterization are simple consequences of the choice of the proposal.  For the low-noise case using simple linear algebra one can verify that
$$
p_n(u_n^{\Delta}(\widetilde{z}_n,\epsilon))
f_n(u_n^{\Delta}(\widetilde{z}_n,x)|u_{n-1}^{\Delta}(\widetilde{z}_{n-1},x)) \propto
\exp\left\{
-\frac{1}{2}z_n^{\top}\widetilde{\Omega}_n^{-1}z_n + \widetilde{\mu}_n^{\top}z_n
\right\}.
$$  
From there one has the optimal proposal and the normalizing constant,  which is (up to a constant) $w_n^{\Delta,\text{opt}}(\widetilde{z}_{n-1})$ can be obtained by completing the square and using Gaussian integrals.  For
the degenerate case we have similarly that
$$
f_n(u_n^{\star}(z_n)|u_{n-1}^{\star}(z_{n-1})) \propto
\exp\left\{
-\frac{1}{2}z_n^{\top}(\Omega_n^{\star})^{-1}z_n + (\mu_n^{\star})^{\top}z_n
\right\}
$$  
and the proof is similar to the low-noise case. The final limit is a direct consequence of \eqref{eq:prf1}
and the dominated convergence theorem.
\end{proof}

\subsection{Verification of Assumption (A\ref{ass:1}) 1.~in the Gaussian case}\label{app:a1_gaussian}

\begin{proof}

We consider the model
\[
Y_n = A_n X_n + \Delta^{1/2}\epsilon_n,
\qquad
X_n = f_n(X_{n-1}) + \nu_n,
\]
where $\epsilon_n \sim \mathcal N(0,\Sigma_n)$ and $\nu_n \sim \mathcal N(0,\Omega_n)$ are independent.
Let $y_n$ be fixed.

Let $x_n^\star \in \mathbb R^{d_x}$ satisfy $A_n x_n^\star = y_n$, and let
$V_n \in \mathbb R^{d_x \times (d_x-d_y)}$ have orthonormal columns spanning $\ker(A_n)$.
Then one can deduce that
\textcolor{blue}{
$$
u_{n}^{\Delta}(\widetilde{z}_n,x) = x_n^\star + V_n z_n - \Delta^{1/2}A_n^\top (A_nA_n^\top)^{-1} \overline{z}_n,
\qquad
u_{n}^{\Delta}(\widetilde{z}_n,\epsilon) = \overline{z}_n.
$$
}


Writing
$m_n := f_n(x_{n-1}) - x_n^\star$,  using the above construction and standard Bayesian calculations the optimal proposal density is
$\tilde q_n^\Delta(\tilde z_n \mid \tilde z_{n-1})$ and has the distribution
\[
\mathcal N_{d_x}(\tilde z_n;\mu_\Delta,\Sigma_\Delta),
\]
where the precision matrix $\Lambda_\Delta = \Sigma_\Delta^{-1}$ is
\textcolor{blue}{
$$
\Lambda_\Delta =
\begin{pmatrix}
V_n^\top\Omega_n^{-1}V_n
&
-\Delta^{1/2}V_n^\top\Omega_n^{-1}A_n^\top (A_nA_n^\top)^{-1}
\\[1mm]
-\Delta^{1/2}(A_nA_n^\top)^{-1}A_n\Omega_n^{-1}V_n
&
\Sigma_n^{-1} + \Delta (A_nA_n^\top)^{-1}A_n\Omega_n^{-1}A_n^\top (A_nA_n^\top)^{-1}
\end{pmatrix},
$$
}
and the linear term $\texttt{L}_{\Delta}$ ($\mu_{\Delta}=\Sigma_{\Delta}\texttt{L}_{\Delta}$) equals
\textcolor{blue}{
$$
\begin{pmatrix}
V_n^\top\Omega_n^{-1}m_n
\\[1mm]
-\Delta^{1/2}(A_nA_n^\top)^{-1}A_n\Omega_n^{-1}m_n
\end{pmatrix}.
$$
}
Since $V_n^\top\Omega_n^{-1}V_n$ and $\Sigma_n^{-1}$ are positive definite,
block matrix inversion yields
\[
\Sigma_\Delta
\;\xrightarrow{\Delta \to 0}\;
\begin{pmatrix}
(V_n^\top\Omega_n^{-1}V_n)^{-1} & 0
\\[1mm]
0 & \Sigma_n
\end{pmatrix}.
\]
Multiplying the limiting covariance with the linear term gives
\[
\mu_\Delta
\;\xrightarrow{\Delta \to 0}\;
\begin{pmatrix}
(V_n^\top\Omega_n^{-1}V_n)^{-1}V_n^\top\Omega_n^{-1}m_n
\\[1mm]
0
\end{pmatrix}
=: \begin{pmatrix} \mu_n^\star \\ 0 \end{pmatrix}.
\]

Hence, as $\Delta \downarrow 0$,  the distribution corresponding to $\tilde q_n^\Delta(z_n,\overline{z}_n \mid \tilde z_{n-1})$ is
\[
\mathcal N(z_n;\mu_n^\star,(V_n^\top\Omega_n^{-1}V_n)^{-1})\,
\mathcal N(\overline{z}_n;0,\Sigma_n).
\]
Writing $q_n^\star(z_n\mid z_{n-1})$ for the first factor and
$p_n(\overline{z}_n)$ for the second, we obtain
\[
\lim_{\Delta\downarrow 0}
\tilde q_n^\Delta(\tilde z_n \mid \tilde z_{n-1})
=
p_n(\overline{z}_n)\, q_n^\star(z_n\mid z_{n-1}),
\]
which verifies Assumption (A\ref{ass:1}) 1.; one can check $\mu_n^\star$ and $(V_n^\top\Omega_n^{-1}V_n)^{-1}$ from Proposition~\ref{prop:prop}.
\end{proof}

\subsection{Verification of Assumption (A\ref{ass:1}) 1.~for conditionally optimal $t$-noise proposal}\label{app:a1_tdist}
\begin{proof}

We consider the nonlinear state-space model used in the numerical example:
\[
Y_n = A_n X_n + \Delta^{1/2}\varepsilon_n,
\qquad
X_n = f(X_{n-1}) + \nu_n,
\]
where $\varepsilon_n \sim \mathcal N(0,\Sigma_n)$ and the transition noise
$\nu_n$ follows a multivariate $t$ distribution
\[
\nu_n \sim t_{d_x}(0,\Omega_n,\vartheta).
\]
The $t$-distributed transition admits the Gaussian scale mixture representation
\[
\nu_n \mid \omega_n \sim \mathcal N\!\left(0,\frac{\Omega_n}{\omega_n}\right),
\qquad
\omega_n \sim \mathrm{Gamma}\!\left(\frac{\vartheta}{2},\frac{\vartheta}{2}\right),
\]
with $\omega_n$ independent across $n$ and independent of all other variables.
Conditionally on $\omega_n$, the state equation becomes
\[
X_n = f(X_{n-1}) + \tilde\nu_n,
\qquad
\tilde\nu_n \sim \mathcal N\!\left(0,\frac{\Omega_n}{\omega_n}\right).
\]

As in Section~\ref{sec:noise_add}, introduce coordinates $\tilde z_n=(z_n,\overline{z}_n)\in\mathbb R^{d_x}$
via
\[
(x_n,\varepsilon_n)
=
\begin{pmatrix}
x_n^\star \\ 0
\end{pmatrix}
+
V_n^\Delta \tilde z_n,
\]
where $A_n x_n^\star=y_n$ and $V_n^\Delta$ is the lifted basis defined in the paper.
This parameterization is independent of $\omega_n$.
Conditionally on $\omega_n$, the proposal used in the numerics is the optimal
Gaussian proposal in the noise parameterization, namely
\[
\tilde q_n^\Delta(\tilde z_n \mid \tilde z_{n-1},\omega_n)
\;\propto\;
p(\varepsilon_n)\,
f(x_n\mid x_{n-1},\omega_n),
\]
where
\[
f(x_n\mid x_{n-1},\omega_n)
=
\mathcal N\!\left(
x_n; f(x_{n-1}),\frac{\Omega_n}{\omega_n}
\right).
\]
For each fixed $\omega_n>0$, this is exactly the Gaussian additive-noise setting
of Appendix~\ref{app:a1_gaussian}, with covariance matrix $\Omega_n/\omega_n$ in place of $\Omega_n$.
Hence, by the Gaussian case already established, we have for each fixed $\omega_n$
\[
\lim_{\Delta\downarrow 0}
\tilde q_n^\Delta(\tilde z_n \mid \tilde z_{n-1},\omega_n)
=
p_n(\overline{z}_n)\,
q_{n,\omega_n}^\star(z_n\mid z_{n-1}),
\]
where $q_{n,\omega_n}^\star$ is the optimal degenerate-noise proposal on the manifold
associated with covariance $\Omega_n/\omega_n$.

The actual proposal used in the algorithm samples $\omega_n$ from its prior
$\mathrm{Gamma}(\vartheta/2,\vartheta/2)$ and then samples
$\tilde z_n$ from the conditional proposal. Thus the marginal proposal density is
\[
\tilde q_n^\Delta(\tilde z_n \mid \tilde z_{n-1})
=
\int
\tilde q_n^\Delta(\tilde z_n \mid \tilde z_{n-1},\omega_n)\,
\pi(\omega_n)\, d\omega_n.
\]
By dominated convergence (the integrand is bounded by an integrable Gaussian
density uniformly in $\Delta$), we may pass the limit inside the integral to obtain
\[
\lim_{\Delta\downarrow 0}
\tilde q_n^\Delta(\tilde z_n \mid \tilde z_{n-1})
=
p_n(\overline{z}_n)
\int q_{n,\omega_n}^\star(z_n\mid z_{n-1})\,\pi(\omega_n)\,d\omega_n.
\]
Define
\[
q_n^\star(z_n\mid z_{n-1})
:=
\int q_{n,\omega_n}^\star(z_n\mid z_{n-1})\,\pi(\omega_n)\,d\omega_n,
\]
which is precisely the degenerate-noise proposal induced by the $t$-distributed
transition.

We have shown that
\[
\lim_{\Delta\downarrow 0}
\tilde q_n^\Delta(\tilde z_n \mid \tilde z_{n-1})
=
p_n(\overline{z}_n)\, q_n^\star(z_n\mid z_{n-1}),
\]
and therefore the conditionally optimal proposal used in the nonlinear
$t$-noise example satisfies Assumption (A\ref{ass:1}).
\end{proof}

\end{document}